\documentclass[pra,twocolumn,preprintnumbers,superscriptaddress,amsmath,amssymb]{revtex4-1}
\usepackage{graphicx}
\usepackage{subfigure}
\usepackage{mathrsfs}
\usepackage{amsfonts}
\usepackage{times}
\usepackage{amsmath}
\usepackage{leftidx}
\usepackage{color}
\usepackage[colorlinks,linkcolor=blue,citecolor=blue]{hyperref}

\newcommand{\Tr}{\operatorname{Tr}} 

\newtheorem{theorem}{Theorem}
\newtheorem{lemma}{Lemma}
\usepackage{bbold}
\usepackage{braket}
\usepackage{mathtools}
\usepackage{physics}

\begin{document}
\title{Error bounds for constrained dynamics in gapped quantum systems: Rigorous results and generalizations}
\author{Zongping Gong}
\affiliation{Department of Physics, University of Tokyo, 7-3-1 Hongo, Bunkyo-ku, Tokyo 113-0033, Japan}
\author{Nobuyuki Yoshioka}
\affiliation{Department of Physics, University of Tokyo, 7-3-1 Hongo, Bunkyo-ku, Tokyo 113-0033, Japan}
\affiliation{Theoretical Quantum Physics Laboratory, RIKEN Cluster for Pioneering Reserach (CPR), Wako-shi, Saitama 351-0198, Japan}
\author{Naoyuki Shibata}
\affiliation{Department of Physics, University of Tokyo, 7-3-1 Hongo, Bunkyo-ku, Tokyo 113-0033, Japan}
\author{Ryusuke Hamazaki}
\affiliation{Department of Physics, University of Tokyo, 7-3-1 Hongo, Bunkyo-ku, Tokyo 113-0033, Japan}
\affiliation{Nonequilibrium Quantum Statistical Mechanics RIKEN Hakubi Research Team, RIKEN Cluster for Pioneering Research (CPR), RIKEN iTHEMS, Wako, Saitama 351-0198, Japan}
\date{\today}

\begin{abstract}
In arXiv:2001.03419 we introduce a universal error bound for constrained unitary dynamics within a well-gapped energy band of an isolated quantum system. Here, we provide the full details on the derivation of the bound. In addition, we generalize the result to isolated quantum many-body systems by employing the local Schrieffer-Wolff transformation, obtaining an error bound that grows polynomially in time. We also generalize the result to Markovian open quantum systems and quantitatively explain the quantum Zeno effect. 
\end{abstract}
\maketitle

\section{Introduction}
In quantum mechanics, the existence of large energy gaps allows us to trace out  the degrees of freedom of irrelevant energy scale~\cite{Sakurai2011}.
Consequently, we can treat a system within a constrained subspace obtained by the projection of the total Hilbert space.
As long as the restricted subspace is energetically well isolated from the remainder of the spectrum, a weak perturbative term that mixes the entire Hilbert space can be treated as an action only on that subspace.
Such approximations have been utilized in various systems, e.g., few-level atoms in quantum optics~\cite{Scully1997} and crystalline materials as few-band systems in condensed matter systems~\cite{Ashcroft1976}.
One of the sophisticated approaches to perform the above approximation is to build a perturbation theory on the basis of the \emph{Schrieffer-Wolff transformation} (SWT)~\cite{Schrieffer1966}, which is  a unitary transformation that gradually block-diagonalizes the Hamiltonian.
First introduced to find an effective theory of the Anderson impurity model~\cite{Schrieffer1966,Paaske2005,Lohneysen2007}, the SWT has widely been applied to many different situations, such as the Bose- and Fermi-Hubbard systems~
\cite{MacDonald1988,Zhang1988,Barthel2009}, quantum dots~\cite{Venturelli2013}, and spins in a cavity~\cite{Heikkila2014,Hwang2015,Burkard2017}. A constrained Hamiltonian obtained by a simple projection can be considered to be associated with the zeroth-order SWT. It is known from the first-order SWT that the error caused by a simple projection is proportional to the strength of the perturbation $V$ and the inverse of the gap $\Delta_0$ between the subspace of interest and the remainder of the spectrum \cite{Bravyi2011}. This means that the effective constrained theory  becomes increasingly accurate when the energy gap becomes large.

Importantly, the projected Hamiltonian is also used to describe the approximated constrained dynamics under the large-gap condition.
For example, quench dynamics in the Bose- and Fermi-Hubbard models are implemented by ultracold atoms in deep optical potentials, which enables us to consider only a ground-state band~\cite{Jaksch1998,Hofstetter2002,Bloch2012b,Gross2020}.
Another notable example is the recent finding of anomalous slow relaxation dynamics of strongly interacting Rydberg systems~\cite{Bernien2017}.
The slow dynamics is effectively described by the so-called PXP model through a projection, which constrains the dynamics in the Hilbert subspace where the adjacent Rydberg excitations are forbidden~\cite{Turner2018}.
The constrained dynamics appears even beyond isolated systems such as a periodically driven setup~\cite{Bukov2016} and a dissipative setup~{\cite{Schafer2014}}.
For the latter case, 
strong dissipation or measurement constrains the dynamics which constitutes the \emph{quantum Zeno effect} \cite{Misra1977,Knight2000,Facchi2002}.
Generalized techniques to apply the SWT to these setups have also been developed \cite{Kessler2012,Weinberg2017}. 

Despite the broad applications of the constrained dynamics, the exact evaluation of the error of the approximation  in the course of the dynamics has remained elusive both for isolated and  open systems. In particular, it stays unclear how the error coming from the perturbation is amplified during the time evolution. Furthermore, another problem appears in quantum many-body systems; the norm of the perturbation $\|V\|$ diverges as the system size increases, so the conventional perturbation analysis becomes inadequate \cite{Bravyi2011}. Solutions to these problems are necessary for the justification of  effective constrained dynamics.

In an accompanying letter~\cite{Gong2020a}, we introduce an error of the expectation value for a normalized observable between the exact dynamics and the effective, constrained dynamics, and rigorously  show that it is universally bounded both in few- and many-body isolated quantum systems.
Such a universal error bound gives a justification of approximated constrained dynamics in generic unitary quantum dynamics, as long as the energy gap is sufficiently large. In this manuscript, we provide the full details of the derivation of the bound with several generalizations.
The core idea is to divide the error between the full quantum dynamics and the constrained one into three terms: an error term for the SWT transformation, a ``Loschmidt-echo" term, and a term for the inverse SWT transformation after time evolution.
Here, the first and third terms come from the generators of  the (inverse) SWT transformations.
The second term is related to the Loschmidt-echo process~\cite{Prosen2006},  in which we go forward in time with  the exact SWT  Hamiltonian and then go backward with the zeroth-order SWT effective Hamiltonian.
We first show that, for a general isolated unitary dynamics, the first and the third terms give a constant error bound, while the  second term gives a bound that grows approximately linearly (see Eq.~\eqref{eq:asymptotic_bound}).
The proof is based on several fundamental inequalities of matrix analysis~\cite{Kato1966,Bhatia1997}.
We then extend our rigorous bound to  locally interacting quantum many-body systems.
Employing the local SWT~\cite{Bravyi2011}, we obtain the error bound for  local observables based on the strength of the \textit{local} perturbation instead of the global  perturbation $\|V\|$.
In this case, the accumulation of the error occurs as a result of the  spreading of initially local operators, which accelerates the error growth from linearly to a power law no faster than $t^{d+1}$ ($d$ is the spatial dimension) due to the Lieb-Robinson bound.
Finally, we investigate a bound for open quantum systems, especially those exhibiting quantum Zeno effect, where strong dissipation or measurement constrains the dynamics.
Using a non-unitary version of the SWT different from Ref.~\cite{Kessler2012},  we prove an error bound similar to the isolated case; the energy scale of dissipation and that of the Hamiltonian take the role of $\Delta_0$ and $\|V\|$, respectively.

The rest of this manuscript is organized as follows.
 In Chapter~\ref{chapter2}, we show two error bounds with a rigorous proof for general isolated quantum systems (see Fig.~\ref{figV1}).
    In Chapter~\ref{chapter3}, we analyze locally interacting quantum many-body systems and show that the error is bounded by a quantity that is again linearly suppressed by the energy gap and grows only polynomially in time.
     In Chapter~\ref{chapter4}, we consider a setup for open quantum systems with strong dissipation or measurement. We show an error bound for the quantum Zeno dynamics and demonstrate its validity with simple models.
    In Chapter~\ref{chapter5}, we summarize our results and discuss future prospects.

\section{Universal error bound for isolated quantum systems}\label{chapter2}
In this section, we briefly review the general setup in Ref.~\cite{Gong2020a} and refine the asymptotic error bound into an exact one. We further discuss an important special case of a single (nondegenerate) isolated eigenenergy.

\subsection{Setup and the main result}
Let us briefly review the setup of gap-induced constrained quantum dynamics. Consider an arbitrarily large quantum system described by a Hamiltonian $H_0$, which has an isolated energy band $\ell_0$ gapped from the remaining by
\begin{equation}
\Delta_0\equiv\min_{\lambda\in\ell_0,\lambda'\in\Lambda_0\backslash\ell_0}|\lambda-\lambda'|,
\end{equation}
where $\Lambda_0$ is the full spectrum of $H_0$. We denote the projector onto this energy band as $P$, which can be expressed as \cite{Kato1966}
\begin{equation}
P=\oint_{C_0}\frac{dz}{2\pi i}R_0(z),\;\;\;\;R_0(z)\equiv\frac{1}{z-H_0},
\label{projres}
\end{equation}
where $R_0(z)$ is called the \emph{resolvent} of $H_0$ and $C_0$ can be an arbitrary closed contour that separates $\ell_0$ from the remaining spectrum. 
This formula will later become useful in Sec.~\ref{SIS}. With an additional coupling $V$ applied, the entire Hamiltonian becomes
\begin{equation}
H=H_0+V.
\label{H0V}
\end{equation}
We aim to provide two universal bounds that are explicitly stated later in Theorem~\ref{Thm:ueb} as Eqs.~\eqref{b1} and \eqref{b2}. While the first bound is unrestricted by the magnitude of the additional coupling, 
the second bound \eqref{b2} is valid when a condition $\|V\| < \frac{1}{2}\Delta_0$, which is sufficient to assure that the full Hamiltonian $H$ is also gapped due to Weyl's perturbation theorem \cite{Bhatia1997}, is satisfied.  Here, $\|\cdot\|$ is the operator norm, i.e., the largest singular value. 

Our objective is to provide an observable-based quantitative bound on the deviation of the constrained dynamics, which is generated by the projected Hamiltonian onto the isolated energy band $H_P\equiv PHP$, from the actual dynamics described by $H$.
Given an observable $O$, which is assumed to be normalized as $\|O\|=1$ without loss of generality, we define the error of the constrained-dynamics approximation as
\begin{equation}
\epsilon(t)\equiv\|P(e^{iHt}Oe^{-iHt}-e^{iH_Pt}Oe^{-iH_Pt})P\|.
\label{ept}
\end{equation}
As is clear from the equivalent variational definition of Eq.~(\ref{ept}): 
\begin{equation}
\epsilon(t)\equiv\max_{\begin{subarray}{l}P|\psi\rangle=|\psi\rangle, \\ \langle\psi|\psi\rangle=1\end{subarray}}|\langle\psi|e^{iHt}Oe^{-iHt}|\psi\rangle-\langle\psi|e^{iH_Pt}Oe^{-iH_Pt}|\psi\rangle|,
\end{equation}
the error tells us in the \emph{worst} case how much the observable's expectation value deviates between the actual dynamics and the constrained one when we start from a state in $\ell_0$.

Our main result is the following theorem:
\begin{theorem}[Universal error bound]
\label{Thm:ueb}
Given an energy gap $\Delta_0$ and a bounded coupling strength $\|V\|$, the error defined in Eq.~(\ref{ept}) is rigorously upper bounded by
\begin{equation}
\epsilon(t)\le \frac{4\|V\|}{\Delta_0}+2(e^{2\frac{\|V\|}{\Delta_0}}-1)\|V\|t,
\label{b1}
\end{equation}
which is a linear function of time. Moreover, when $\|V\|<\frac{1}{2}\Delta_0$, we have another rigorous bound
\begin{equation}
\epsilon(t)\le \frac{4\|V\|}{\Delta_0-2\|V\|}+2f\left(\frac{2\|V\|}{\Delta_0-2\|V\|}\right)\|V\|t,
\label{b2}
\end{equation}
which is again linear in $t$ and the function $f$ is given by 
\begin{equation}
f(x)=\frac{(x-1)e^x+1}{x}.
\end{equation}
\end{theorem}

\begin{figure}
\begin{center}
       \includegraphics[width=6.5cm, clip]{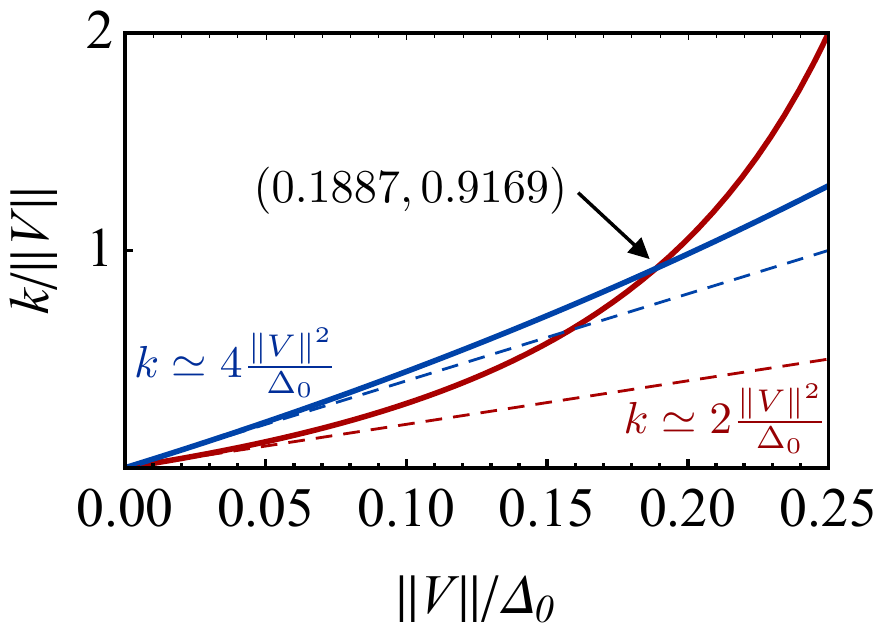}
       \end{center}
   \caption{ Slopes $k$ of Eq.~(\ref{b1}) (blue curve) and Eq.~(\ref{b2}) (red curve) for various $\frac{\|V\|}{\Delta_0}$. The latter is smaller when $\|V\|<0.1887\Delta_0$. Asymptotically, both slopes scale linear with $\frac{\|V\|^2}{\Delta_0}$ (dashed lines) with prefactors $4$ and $2$, respectively.}
   \label{figV1}
\end{figure}

Before diving into the detailed proof, let us first make a few remarks on Theorem~\ref{Thm:ueb}. First, let us compare the time slopes in Eqs.~(\ref{b1}) and (\ref{b2}). For a sufficiently long time, a smaller slope should give a tighter bound. As shown in Fig.~\ref{figV1}, as long as $\|V\|<0.1887\Delta_0$, the slope in Eq.~(\ref{b2}) is smaller. Second, in the large gap regime $\Delta_0\gg\|V\|$, we can easily check that the slope in Eq.~(\ref{b1}) asymptotically approaches $\frac{4\|V\|^2}{\Delta_0}$ while that in Eq.~(\ref{b2}) approaches $\frac{2\|V\|^2}{\Delta_0}$. Also, the intercept in Eq.~(\ref{b2}) deviates from that in Eq.~(\ref{b1}) by a quantity of the order of $\frac{\|V\|^2}{\Delta_0^2}$. Therefore, Eq.~(\ref{b2}) turns out to be tighter in this large-gap regime and the leading term (order $\frac{\|V\|}{\Delta_0}$) indeed reproduces the asymptotic bound in Ref.~\cite{Gong2020a}:
\begin{equation}
\epsilon(t)\lesssim\frac{4\|V\|}{\Delta_0}+\frac{2\|V\|^2}{\Delta_0}t,\label{eq:asymptotic_bound}
\end{equation}
where “$\lesssim$” means that there could be a tiny violation up to $\order{\frac{\|V\|^2}{\Delta_0^2}}$.

\subsection{Proof of the main result}
\label{pmr}
\begin{figure}
\begin{center}
       \includegraphics[width=8.5cm, clip]{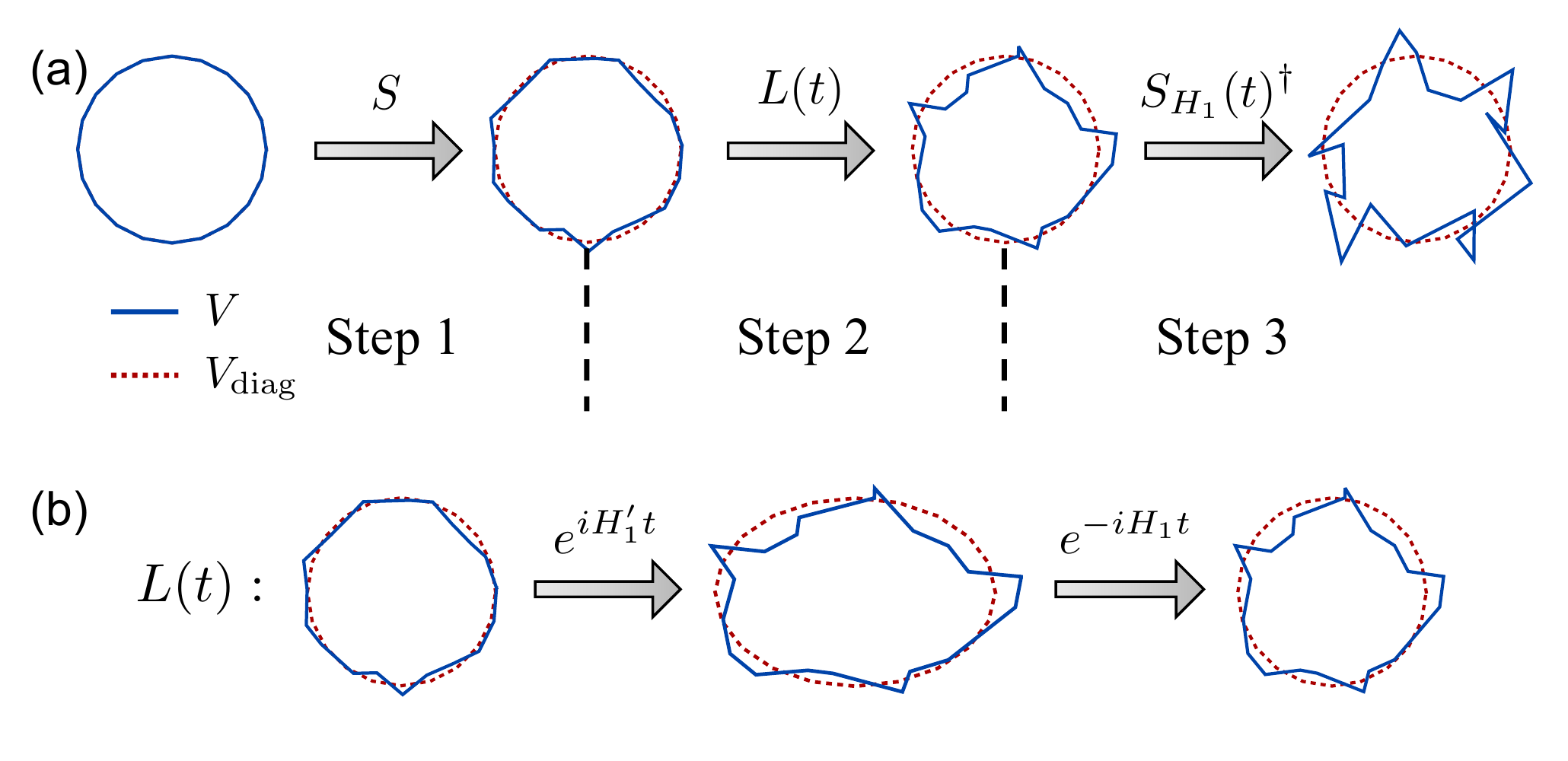}
       \end{center}
   \caption{(a) Schematic illustration of the three steps of error production. The blue solid and red dashed shapes correspond to the operator evolution in Eq.~(\ref{dec}) for the full coupling $V$ and the block-diagonal component $V_{\rm diag}$, respectively, and their difference represents the error. Here $S$ is the unitary SWT generated by an anti-Hermitian operator $T$ determined by Eq.~(\ref{H0T}), $L(t)$ is the Loschmidt-echo operator (\ref{Lt}) and $S_{H_1}(t)$ is the SWT in the interacting picture (\ref{SH1t}). (b) Explicit expression of the Loschmidt echo operator $L(t)=e^{-iH_1t}e^{iH'_1t}$, where $H_1'\equiv H_1+V'$ (see Eq.~(\ref{H1p})).}
   \label{figV2}
\end{figure}

The general idea of the proof has been mentioned in Ref.~\cite{Gong2020a}. The crucial point is to rewrite Eq.~(\ref{ept}) following the definition of the SWT as follows:
\begin{equation}
\epsilon(t)=\|P[S_{H_1}(t)^\dag L(t)SOS^\dag L(t)^\dag S_{H_1}(t)-O]P\|.
\label{dec}
\end{equation}
See Fig.~\ref{figV2} for a schematic illustration of such a decomposition. In order to first obtain Eq.~(\ref{b1}), here we require the anti-Hermitian generator $T$ in the unitary SWT $S=e^T$ to satisfy 
\begin{equation}
[H_0,T]=PVQ+QVP\equiv V_{\rm off}.
\label{H0T}
\end{equation}
This is slightly different from the derivation in Ref. \cite{Gong2020a}, where $H_0$ in Eq.~(\ref{H0T}) is replaced by $H_1$ (see the definition in Eq.~(\ref{H1})). Formally, the other operators in Eq.~(\ref{dec}) are defined in a similar way as in Ref.~\cite{Gong2020a}: 
\begin{equation}
L(t)=e^{-iH_1t}e^{iH_1't} 
\label{Lt}
\end{equation}
is the Loschmidt-echo operator and 
\begin{equation}
S_{H_1}(t)=e^{-iH_1 t} e^T e^{iH_1 t} =e^{e^{-iH_1t}Te^{iH_1t}} 
\label{SH1t}
\end{equation}
is the SWT in the interacting picture with respect to $H_1$. Here
\begin{equation}
H_1\equiv H_P+H_Q=H_0+V_{\rm diag},
\label{H1}
\end{equation}
where $Q=1-P$, $H_Q\equiv QHQ$ and $V_{\rm diag}\equiv PVP+QVQ=V-V_{\rm off}$ is the block-diagonal component of $V$. 
\begin{equation}
H_1'\equiv SHS^\dag=H_1+V'
\label{H1p}
\end{equation}
is related to $H$ via the SWT. However, since $H_0$ instead of $H_1$ is used in Eq.~(\ref{H0T}) to determine $T$, the expression of $V'$ differs slightly from Eq.~(10) in Ref.~\cite{Gong2020a} unless $V=V_{\rm off}$:
\begin{equation}
V'=\sum^\infty_{n=1} \frac{1}{n!}{\rm ad}^n_T\left(V-\frac{1}{n+1}V_{\rm off}\right),
\label{Vp}
\end{equation}
where ${\rm ad}_T(\;\cdot\;)\equiv[T,\;\cdot\;]$. With all the quantities in Eq.~(\ref{dec}) clarified above, we can straightforwardly verify its equivalence to Eq.~(\ref{ept}): % and (\ref{dec}):
\begin{equation}
\begin{split}
\epsilon(t)&=\|P(e^{-iH_1t}S^\dag e^{iH'_1t}SOS^\dag  e^{-iH'_1t}Se^{iH_1t}-O)P\| \\
&=\|P(e^{iS^\dag H'_1St}O e^{-iS^\dag H'_1St}-e^{iH_1t}Oe^{-iH_1t})P\| \\
&=\|P(e^{iHt}Oe^{-iHt}-e^{iH_Pt}Oe^{-iH_Pt})P\|,
\end{split}
\end{equation}
where we have also used the unitary invariance of the operator norm and the identity $e^{iH_1t}P=Pe^{iH_1t}=Pe^{iH_Pt}$, which arises from $PH_1=H_1P=H_P$.

To proceed, we upper bound Eq.~(\ref{dec}) by three terms
\begin{equation}
\epsilon(t)\le\|[S,O]\|+\|[L(t),O]\|+\|[S_{H_1}(t),O]\|,
\label{decineq}
\end{equation}
where $\|P(\;\cdot\;)P\|\le \|\cdot\|$ and the inequality
\begin{equation}
\left\|\left[\prod_\alpha U_\alpha,O\right]\right\|\le\sum_\alpha\|[U_\alpha,O]\|
\label{UaO}
\end{equation}
for unitaries $U_\alpha$'s has been used. To prove Eq.~(\ref{UaO}), it is sufficient to iteratively apply the inequality for two unitaries:
\begin{equation}
\begin{split}
\|[UU',O]\|&=\|U[U',O]+[U,O]U'\| \\
&\le\|[U,O]\|+\|[U',O]\|,
\end{split}
\label{U1U2}
\end{equation}
where we have used the unitary invariance of operator norm. By iteratively applying Eq.~(\ref{U1U2}), we mean that we can first choose $U$ and $U'$ in Eq.~(\ref{U1U2}) to be $U_1$ and $\prod_{\alpha>1}U_\alpha$ in Eq.~(\ref{UaO}), respectively, to obtain $\|[\prod_\alpha U_\alpha,O]\|\le\|[U_1,O]\|+\|[\prod_{\alpha>1} U_\alpha,O]\|$. Then we choose $U=U_2$ and $U'=\prod_{\alpha>2}U_\alpha$ to obtain $\|[\prod_\alpha U_\alpha,O]\|\le\|[U_1,O]\|+\|[U_2,O]\|+\|[\prod_{\alpha>2} U_\alpha,O]\|$, etc. Moreover, the norm of the commutator between an operator and a unitary can be bounded in terms of the anti-Hermitian generator by rewriting the commutator using the integral:
\begin{equation}
\begin{split}
&\|[e^T,O]\|=\|e^TOe^{-T}-O\| \\
=&\left\|\int^1_0d\lambda e^{\lambda T}[T,O]e^{-\lambda T}\right\| \\
\le&\int^1_0d\lambda \|e^{\lambda T}[T,O]e^{-\lambda T}\| \\
=&\|[T,O]\|\le 2\|T\|.
\end{split}
\label{eTO}
\end{equation}
This result applies directly to the first and the third terms (where $T$ is replaced by $e^{-iH_1t}Te^{iH_1t}$, whose norm is the same as $\|T\|$) in Eq.~(\ref{decineq}), while for the middle term we should use a time-dependent version of Eq.~(\ref{eTO}):
\begin{equation}
\begin{split}
&\|[L(t),O]\|=\|L(t)^\dag OL(t)-O\| \\ 
=&\left\|\int^t_0dt'L(t')^\dag[e^{-iH_1t'}V'e^{iH_1t'},O]L(t')\right\| \\
\le &\int^t_0dt' \|[e^{-iH_1t'}V'e^{iH_1t'},O]\| \\
\le &2\int^t_0dt'\|e^{-iH_1t'}V'e^{iH_1t'}\|=2\|V'\|t,
\end{split}
\label{LtO}
\end{equation}
Combining Eqs.~(\ref{eTO}) and (\ref{LtO}) with Eq.~(\ref{decineq}), we obtain
\begin{equation}
\epsilon(t)\le4\|T\|+2\|V'\|t.
\label{eptTVp}
\end{equation}

The remaining problem is how we can bound $\|T\|$ and $\|V'\|$ in terms of the energy gap $\Delta_0$ and the coupling strength $\|V\|$. We first recall that $T$ is determined by Eq.~(\ref{H0T}). While the solution is not unique, we can further impose the constraint that $T$ is off-block-diagonal, so that one of the off-block-diagonal components $T_{PQ}\equiv PTQ$ can uniquely be determined from the following \emph{Sylvester equation} \cite{Sylvester1884}:
\begin{equation}
H_{0P}T_{PQ}-T_{PQ}H_{0Q}=V_{PQ}\equiv PVQ,
\label{TPQ}
\end{equation}
where $H_{0P}\equiv PH_0P$ and $H_{0Q}\equiv QH_0Q$. Note that the singular value spectrum of $T$ coincides with the positive half of the spectrum of $iT$, and in particular
\begin{equation}
\|T\|=\|T_{PQ}\|.
\label{TTPQ}
\end{equation}
To see this, we only have to note that $T^2=PTQTP+ QTPTQ$ is block-diagonalized. This implies $\|T\|^2=\max\{\|PTQTP\|,\|QTPTQ\|\}=\|QTP\|^2=\|PTQ\|^2$ (due to $\|A\|^2=\|A^\dag A\|$), which gives the desired result. By assumption, it is possible to find $E_0\in[\min \Lambda_0,\max\Lambda_0]$ ($\Lambda_0$ is the spectrum of $H_0$) and $r_0\ge0$ such that $\ell_0$ is covered by $[E_0-r_0,E_0+r_0]$ and $\Lambda_0\backslash\ell_0$ is covered by $(-\infty,E_0-r_0-\Delta_0]\cup[E_0+r_0+\Delta_0,\infty)$. Therefore, we can iteratively apply 
\begin{equation}
T_{PQ}=[(H_{0P}-E_0)T_{PQ}-V_{PQ}](H_{0Q}-E_0)^{-1}
\end{equation}
to obtain a formal solution to Eq.~(\ref{TPQ}) as
\begin{equation}
T_{PQ}=-\sum^\infty_{n=0}(H_{0P}-E_0)^nV_{PQ}(H_{0Q}-E_0)^{-(n+1)}.
\end{equation}
The convergence of this series can be seen from the (no slower than) exponential decay in the norm of the $n$th term:
\begin{equation}
\begin{split}
&\|(H_{0P}-E_0)^nV_{PQ}(H_{0Q}-E_0)^{-(n+1)}\| \\
\le&\|H_{0P}-E_0\|^n\|H_{0Q}-E_0\|^{-(n+1)}\|V_{PQ}\| \\
\le&\frac{r_0^n}{(r_0+\Delta_0)^{n+1}}\|V\|.
\end{split}
\label{series}
\end{equation}
Moreover, after summing these inequalities up and using Eq.~(\ref{TTPQ}), we obtain \cite{Bhatia1997}
\begin{equation}
\|T\|\le\sum^\infty_{n=0}\frac{r_0^n\|V\|}{(r_0+\Delta_0)^{n+1}}=\frac{\|V\|}{\Delta_0}.
\label{Tb}
\end{equation}
Such a bound on $\|T\|$ allows us to bound $\|V'\|$ as
\begin{equation}
\begin{split}
\|V'\|\le&\sum^\infty_{n=1} \frac{1}{n!}\left\|{\rm ad}_T^n\left(V-\frac{1}{n+1}V_{\rm off}\right)\right\| \\
\le&\sum^{\infty}_{n=1}\frac{(2\|T\|)^n}{n!}\|V\|=(e^{2\|T\|}-1)\|V\| \\
\le&(e^{\frac{2\|V\|}{\Delta_0}}-1)\|V\|,
\end{split}
\label{Vpb}
\end{equation}
where we have used $\|{\rm ad}_T(\;\cdot\;)\|\le 2\|T\|\|\cdot\|$ and
%\begin{equation}
%\begin{split}
$\|V-\frac{1}{n+1}V_{\rm off}\|=\|\frac{n}{n+1}V+\frac{1}{n+1}V_{\rm diag}\|
\le\frac{n}{n+1}\|V\|+\frac{1}{n+1}\|V_{\rm diag}\|\le\|V\|$.
%\end{split}
%\end{equation}
The first main result Eq.~(\ref{b1}) follows the combination of Eqs.~(\ref{Tb}), (\ref{Vpb}) and (\ref{eptTVp}).

So far, we have solely assumed the gap $\Delta_0>0$ to derive the  bound Eq.~\eqref{b1} based on the  decomposition of the error~\eqref{ept} into three terms: the SWT and the time-evolved SWT that contribute to the constant term and the Loschmidt-echo operator which is the source of the time-linear term. The generator of the SWT is defined such that the commutation relation with $H_0$ \eqref{H0T} is satisfied.

To derive the second bound (\ref{b2}), we should choose \emph{another} $T$ to satisfy 
\begin{equation}
[T,H_1]=-V_{\rm off},
\label{neuT}
\end{equation}
which coincides with the choice in Ref.~\cite{Gong2020a}.
According to Weyl's perturbation theorem \cite{Weyl1912}, the spectral shift of $H_1=H_0+V_{\rm diag}$ compared with $H_0$ is rigorously bounded by $\|V_{\rm diag}\|\le \|V\|$. Accordingly, the energy gap of $H_1$ satisfies
\begin{equation}
\Delta_1\ge\Delta_0- 2\|V\|,
\end{equation}
implying that the perturbed energy band $\ell$ stays isolated if $\|V\|<\frac{1}{2}\Delta_0$. In this case,  the energy gap in Eq.~(\ref{Tb}) is replaced with $\Delta_1$ owing to the change in the definition of $T$, whose norm is now bounded by
% In this case,  Eq.~(\ref{Tb}) becomes
\begin{equation}
\|T\|\le \frac{\|V\|}{\Delta_1}\le  \frac{\|V\|}{\Delta_0-2\|V\|}.
\label{Tb2}
\end{equation}
We can thus bound the norm of 
\begin{equation}
V'=\sum_{n=1}^{\infty}\frac{1}{n!} {\rm ad}_T^n\left(\frac{n}{n+1}V_{\rm off}\right), 
\end{equation}
which is also replaced due to the modification in the definition (\ref{neuT}) of $T$,
%given by Eq.~(\ref{Vp}) 
as follows:
\begin{equation}
\begin{split}
\|V'\|&\le\sum^\infty_{n=1}\frac{n(2\|T\|)^n}{(n+1)!}\|V_{\rm off}\|\le f(2\|T\|)\|V\| \\
&\le f\left(\frac{2\|V\|}{\Delta_0-2\|V\|}\right)\|V\|,
\end{split}
\label{Vpb2}
\end{equation}
where we have used $\|V_{\rm off}\|=\|PVQ\|\le \|V\|$ (cf. Eq.~(\ref{TTPQ})) and
\begin{equation}
\begin{split}
\sum^\infty_{n=1}\frac{nx^n}{(n+1)!}&=\sum^n_{n=1}\frac{x^n}{n!}-x^{-1}\sum^\infty_{n=1}\frac{x^{n+1}}{(n+1)!} \\
&=e^x-1-x^{-1}(e^x-x-1)=f(x),
\end{split}
\end{equation}
which is monotonically increasing on $\mathbb{R}^+$. 

Finally, we mention that the results can readily be generalized to the case of multiple energy bands. The only price which we have to pay for this is to multiply $\frac{\pi}{2}$ to each $\frac{\|V\|}{\Delta_0}$ or $\frac{\|V\|}{\Delta_0-2\|V\|}$ in the bounds. This is because the solution $X$ of the Sylvester equation $AX-XB=Y$ with Hermitian $A$ and $B$ always satisfies $\|X\|\le\frac{\pi}{2}\frac{\|Y\|}{\Delta}$ \cite{Bhatia1997}, where $\Delta\equiv\min_{\lambda\in\Lambda_A,\lambda'\in\Lambda_B}|\lambda-\lambda'|$ is the distance between the spectra of $A$ and $B$ (denoted as $\Lambda_A$ and $\Lambda_B$). However, it is far from clear whether these bounds can (partially) be saturated in the worst cases.

\subsection{Case of a single isolated state}
\label{SIS}
While the linear growth of the error and even the asymptotic saturation of the slope in the universal error bound can be already achieved by a two-level energy band \cite{Gong2020a}, things become qualitatively different when the isolated energy band consists of only a single state $|\psi\rangle$. In this case, the constrained dynamics is simply ``no dynamics" and we can actually prove that the error for a normalized observable can never exceed an $\mathcal{O}(\frac{\|V\|}{\Delta_0})$ constant even in the infinite-time limit, and thus a time-linear term is unnecessary in the error bound. This result can intuitively be understood from the standard perturbation theory for static eigenstates, which tells us that $|\psi\rangle$ should have a large overlap with the corresponding eigenvector of $H$ when the perturbation is sufficiently small compared to the gap~\cite{Sakurai2011}. Accordingly, the evolved state $e^{-iHt}|\psi\rangle$ also has a large overlap with $|\psi \rangle$ and any observable should stay almost unchanged. In the following, we translate this argument into a rigorous proof.

Since the projector $P=|\psi\rangle\langle\psi|$ is of rank one, we find $e^{-iH_Pt}|\psi\rangle=e^{-i\langle\psi|H|\psi\rangle t}|\psi\rangle$ and hence the error can be rewritten as
\begin{equation}
\begin{split}
\epsilon(t)&=|\langle\psi|e^{iHt}Oe^{-iHt}|\psi\rangle - \langle\psi|O|\psi\rangle| \\
&=|\Tr[O(e^{-iHt}Pe^{iHt}-P)]|,
\end{split}
\end{equation}
for which we can apply the H\"older inequality $|\Tr[A^\dag B]|\le \|A\|_1\|B\|$ ($\|A\|_1\equiv\Tr\sqrt{A^\dag A}$ is the Schatten-$1$ norm) \cite{Baumgartner2011} to obtain
\begin{equation}
\epsilon(t)\le\|e^{-iHt}Pe^{iHt}-P\|_1=2\|e^{-iHt}Pe^{iHt}-P\|,
\label{etr1}
\end{equation}
where $\|O\|=1$ has been used. The equality in Eq.~(\ref{etr1}) is due to the fact that $e^{-iHt}Pe^{iHt}-P$ is of rank $2$ (unless $P=e^{-iHt}Pe^{iHt}$, in which case the equality stays valid) and traceless, so there are two eigenvalues which are the opposite of each other. Provided that $\|V\|<\frac{1}{2}\Delta_0$ so that the isolated eigenenergy is ensured to stay isolated by Weyl's perturbation theorem \cite{Weyl1912}, we have a well-defined projector $P'=|\psi'\rangle\langle\psi'|$ onto the perturbed eigenstate $|\psi'\rangle$ of $H$. By definition, $[P',H]=0$ and thus
\begin{equation}
\begin{split}
\epsilon(t)&\le 2\|e^{-iHt}(P-P')e^{iHt}-(P-P')\| \\
&\le 4\|P-P'\|.
\end{split}
\label{etnP}
\end{equation}
So far, we have bounded the error for dynamics by that for static eigenprojectors, and it suffices to show that $\|P-P'\|$ is a small quantity of the order of $\frac{\|V\|}{\Delta_0}$.

\begin{figure}
\begin{center}
       \includegraphics[width=8cm, clip]{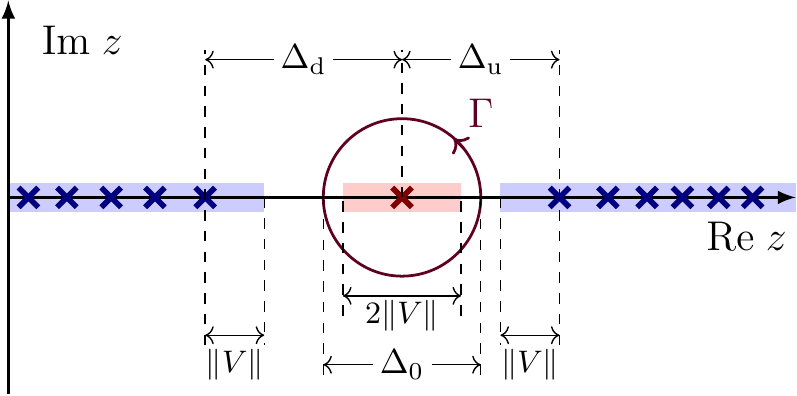}
       \end{center}
   \caption{ Poles (marked as ``$\times$") of the resolvent $R_0(z)=(z-H_0)^{-1}$ forming the spectrum of $H_0$. An isolated eigenenergy (red) is separated from the remaining (blue) by a gap $\Delta_0=\min\{\Delta_{\rm d},\Delta_{\rm u}\}$. When $H_0$ is changed into $H=H_0+V$, the eigenvalues cannot shift by more than $\|V\|$ \cite{Bhatia1997} and should thus always be covered by the shaded regions. Provided that $\|V\|<\frac{1}{2}\Delta_0$, the circular contour $\Gamma$ with diameter $\Delta_0$ separates the isolated eigenenergies of both $H_0$ and $H$ from the remaining.}
   \label{figV3}
\end{figure}

To estimate the shift of the projector, we apply the perturbative formula \cite{Kato1966}
\begin{equation}
P'-P=\sum^\infty_{n=1}\oint_\Gamma \frac{dz}{2\pi i} R_0(z)[VR_0(z)]^n,
\end{equation}
where $R_0(z)$ is the resolvent of $H_0$ given in Eq.~(\ref{projres}) and $\Gamma$ can be an arbitrary closed contour that separates the isolated eigenvalues in $H_0$ and $H$ from the remaining. The existence of $\Gamma$ is ensured when $\|V\|<\frac{1}{2}\Delta_0$, and, in particular, we can choose it to be a circle centered at the isolated eigenenergy of $H_0$ and with diameter $\Delta_0$ (see Fig.~\ref{figV3}). For such a particular choice, we have
\begin{equation}
\begin{split}
\|P-P'\|&\le \frac{l_\Gamma}{2\pi}\sum^\infty_{n=1} \|V\|^n (\max_{z\in\Gamma}\|R_0(z)\|)^{n+1} \\
&=\sum^\infty_{n=1} (2\Delta_0^{-1}\|V\|)^n=\frac{2\|V\|}{\Delta_0 - 2\|V\|},
\end{split}
\label{PmPp}
\end{equation}
where $l_\Gamma=\pi\Delta_0$ is the length of $\Gamma$ and $\max_{z\in\Gamma}\|R_0(z)\|=2\Delta_0^{-1}$. Substituting Eq.~(\ref{PmPp}) into Eq.~(\ref{etnP}), we obtain
\begin{equation}
\epsilon(t)\le\frac{8\|V\|}{\Delta_0-2\|V\|},
\label{constb}
\end{equation}
implying that at any time the error is rigorously bounded by a constant of the order of $\frac{\|V\|}{\Delta_0}$. While Eq.~(\ref{constb}) already establishes the desired result, it stays an open problem whether the constant bound is optimal, i.e., whether it can be saturated in the worst case.

\section{Generalization to quantum many-body systems}\label{chapter3}
Let us turn to the case of isolated quantum systems with a macroscopic number of degrees of freedom. As already mentioned in Ref.~\cite{Gong2020a}, for such quantum many-body systems we typically encounter the situation in which $V$ is a sum of local operators, so its norm diverges %linearly 
in the thermodynamic limit and we cannot apply the error bound in Eq.~(\ref{b1}), let alone Eqs.~(\ref{b2}) and (\ref{eq:asymptotic_bound}). Nevertheless, by making fully use of locality, we can still derive a bound for \emph{local} observables, which is of the order of $\Delta_0^{-1}$ and
grows no faster than polynomially in time. This is achieved by a combination of the local SWT \cite{Bravyi2011,Datta1996} and the Lieb-Robinson bound \cite{Lieb1972,Nachtergaele2006,Bravyi2006}. We note that similar ideas have been applied to prove the quantum adiabatic theorem for many-body systems \cite{Bachmann2017}, which states that while the adiabatic (instantaneous ground state) approximation generally breaks down globally, it stays valid locally.

\subsection{Local Schrieffer-Wolff transformation for quantum many-body Hamiltonians}
We consider a quantum many-body Hamiltonian defined on a $d$-dimensional lattice $\Lambda$, where each site is associated with a finite dimensional local Hilbert space. The many-body Hamiltonian takes the form of Eq.~(\ref{H0V}), where $H_0$ and $V$ are local, i.e., they can be written as a sum of Hermitian operators supported on finite regions, whose norms are uniformly bounded. Moreover, $H_0$ is assumed to be \emph{commuting}, \emph{frustration-free} \cite{Spyridon2013} and admit an \emph{exactly} degenerate low-energy manifold $\mathcal{H}_P$. By commuting, we mean that all the local terms in $H_0$ commute with each other.
%all the eigenstates of $H_0$ are simply \emph{Fock states}, which take the form $\otimes^{|\Lambda|}_{j=1}|n_j\rangle$ ($|\Lambda|$: total number of sites) with the $j$th-site state $|n_j\rangle$ being one of the states in a \emph{fixed} orthonormal basis of the local Hilbert space. 
By frustration-free, we mean that each eigenstate %(Fock state) 
in $\mathcal{H}_P$ not only has the lowest global energy by definition, but also minimizes the local energy everywhere. We emphasize that assuming the exact degeneracy is not a severe limitation of our theory, since we can always recast the degeneracy-lifting perturbations into $V$.

Without loss of generality, we assume $\mathcal{H}_P$ to be the \emph{kernel}, i.e., the zero-energy eigenspace of $H_0$, so that we can decompose $H_0$ into
\begin{equation}
H_0=\sum_j H_{0j},
\end{equation}
where $H_{0j}$ is supported on a finite region $R_j$, which contains the $j$th site,  and positive-semidefinite, and admits a nonempty kernel, onto which the projector is denoted as $P_j$. Recalling that $H_0$ is assumed to be commuting and frustrated-free, %classical, 
we have
\begin{equation}
[H_{0j},H_{0j'}]=[P_j,P_{j'}]=0,
\label{HHPP}
\end{equation}
and the projector onto the global kernel $\mathcal{H}_P$ is given by 
\begin{equation}
P=P_{\Lambda}\equiv \prod_{j\in\Lambda} P_j, 
\end{equation}
where we define for all $ A\subseteq\Lambda$
%\begin{equation}
%P=\prod_{j\in\Lambda} P_j.
%\end{equation}
%where each $P_j$ supported on a finite region $R_j$ containing the $j$th site. By classical, we mean that the many-body eigenstates of $H_0$ are simply Fock states and, in particular, $[P_j,P_{j'}]=0$ for all $ j,j'$. Without loss of generality, we assume the eigenenergy of any state in $\mathcal{H}_P$ to be zero. then obviously $P=P_\Lambda$.
\begin{equation}
P_A\equiv\prod_{R_j\cap A\neq\emptyset} P_j. 
\label{PA}
\end{equation}
We further require $H_0$ to have an energy gap $\Delta_0$, then $\Delta_0$ uniformly lower bounds the gap of any locally truncated Hamiltonian:
\begin{equation}
H_{0A}\equiv \sum_{R_j\cap A\neq\emptyset} H_{0j}. 
\end{equation}

The many-body perturbation $V$ can explicitly be written as 
\begin{equation}
V=\sum_{A\subseteq\Lambda} V_A,
\end{equation}
where $V_A$ is Hermitian and supported on $A$. Since $V$ is also assumed to be local, we have $V_A=0$ whenever $A$ is not connected or the \emph{volume} of $A$, which is defined as the number of sites and is denoted as $|A|$, exceeds some threshold. Such a (strict) locality implies that, even in the thermodynamic limit $|\Lambda|\to\infty$ ($|\Lambda|$: total number of sites), for all $\mu\ge0$ we can define
\begin{equation}
\|V\|_\mu\equiv\max_{j\in\Lambda}\sum_{A\ni j}\|V_A\|e^{\mu l_A}<\infty,
\label{qloc}
\end{equation}
where $l_A\equiv\max_{x,y\in A}{\rm dist}(x,y)$ denotes the \emph{diameter} of $A$. %\footnote{Our definition of the diameter differs from that in Ref.~\cite{Hastings2010} by ``$+1$''. Such a modification allows us to bound the volume of a connected $d$-dimensional region $A$ by $C_d l_A^d$ for some constant $C_d\sim\mathcal{O}(1)$ even if $|A|=1$.}. 
Here ${\rm dist}$ is defined on a general graph as the minimal number of edges that connect two vertices \cite{Hastings2010}. For example, in a cubic lattice $\Lambda\subseteq\mathbb{Z}^d$, ${\rm dist}(x,y)\equiv\sum^d_{\alpha=1}|x_\alpha-y_\alpha|$ with $x_\alpha\in\mathbb{Z}$ ($y_\alpha\in\mathbb{Z}$) being the $\alpha$th component of $x$ ($y$). The quantity defined in Eq.~(\ref{qloc}) measures the largest local interaction strength in $V$. Obviously, we can get rid of $\max_{j\in\Lambda}$ and set $j=0$ or any other site if the system is translation-invariant, but our analysis does not require this and thus applies equally to disordered systems. In the following, we will also encounter \emph{quasi-local} interactions, which may have nonzero but exponentially small (in terms of $l_A$) $\|V_A\|$ for a large $A$ and the interaction norm $\|V\|_\mu$ in Eq.~(\ref{qloc}) is well-defined only for a sufficiently small $\mu$. In particular, 
\begin{equation}
\|V\|_\star\equiv\|V\|_{\mu=0}=\max_{j\in\Lambda}\sum_{A\ni j}\|V_A\| 
\end{equation}
is always well-defined in this case.

We are now well-prepared to introduce the \emph{local} SWT. This is to be contrasted to the \emph{global} SWT, which is nothing but the conventional SWT (used in Sec.~\ref{chapter2}) applied to a many-body Hamiltonian. In the latter case, the generator of the SWT is usually highly non-local, while that in the former case is constructed to be local. To this end, instead of globally block diagonalizing the many-body Hamiltonian, we require the SWT to block diagonalize each local term. Provided that the gap $\Delta_0$ is large enough, up to the first order, the local SWT $S=e^T$ is chosen such that \cite{Bravyi2011}
\begin{equation}
%H_S\equiv 
SHS^\dag=H_0+\sum_{A\subseteq\Lambda}\mathscr{D}_A(V_A)+V'\equiv H_1+V',
\label{HS}
\end{equation}
where $\mathscr{D}_A(V_A)\equiv P_AV_AP_A+Q_AV_AQ_A$ with $P_A$ being defined in Eq.~(\ref{PA}), $Q_A\equiv 1-P_A$, $V'$ is a quasi-local interaction with bounded $\|V'\|_\mu$ for a sufficiently small $\mu$ (see Appendix~\ref{bVp}), and $T$ is an anti-Hermitian local interaction determined by
\begin{equation}
T=\sum_{A\subseteq\Lambda}\mathscr{L}_A(V_A).
\label{Tloc}
\end{equation}
Here the superoperator $\mathscr{L}_A$ is defined as
\begin{equation}
\begin{split}
\mathscr{L}_A(V_A)&\equiv\int^\infty_0 d\tau (e^{-\tau H_{0A}}Q_AV_AP_A-{\rm H.c.}) \\
&=(Q_A H_{0A}Q_A)^{-1}Q_AV_AP_A-{\rm H.c.},
\end{split}
\end{equation}
which satisfies
\begin{equation}
\begin{split}
&\;\;\;[H_0,\mathscr{L}_A(V_A)]=[H_{0A},\mathscr{L}_A(V_A)] \\
&=H_{0A}(Q_A H_{0A}Q_A)^{-1}Q_AV_AP_A+{\rm H.c.}\\
&=Q_AV_AP_A+{\rm H.c.}=\mathscr{O}_A(V_A)\equiv V_A-\mathscr{D}_A(V_A),
\end{split}
\end{equation}
where we have used $P_AH_{0A}=H_{0A}P_A=0$, Eq.~(\ref{HHPP}) and $[H_{0j},V_A]=0$ for all $j\in\Lambda$ satisfying $R_j\cap A=\emptyset$. We note that the support of a local term $\mathscr{L}_A(V_A)$ in $T$ is generally not $A$ but instead 
\begin{equation}
R_A\equiv \bigcup_{R_j\cap A\neq\emptyset}R_j.
\label{RA}
\end{equation}
Also, using Eq.~(\ref{TTPQ}) and the fact that the gap of $H_{0A}$ is lower bounded by $\Delta_0$, we have
\begin{equation}
\|\mathscr{L}_A(V_A)\|=\|(Q_A H_{0A}Q_A)^{-1}Q_AV_AP_A\|\le\frac{\|V_A\|}{\Delta_0}.
\label{LAVA}
\end{equation}
While the local-SWT generator $T$ has a divergent norm in the thermodynamic limit, we can show from Eq.~(\ref{LAVA}) that the \emph{local} interaction strength $\|T\|_\star\equiv\|T\|_{\mu=0}$ is finite and linearly suppressed by the energy gap $\Delta_0$: 
\begin{equation}
\begin{split}
\|T\|_\star&=\max_{j\in\Lambda}\sum_{R_A\ni j}\|\mathscr{L}_A(V_A)\|\\
&\le\max_{j\in\Lambda}\sum_{A\cap R_{j'}\neq\emptyset,\exists R_{j'}\ni j}\frac{\|V_A\|}{\Delta_0} \\
&\le\max_{j\in\Lambda}\sum_{x\in R_{j'},\exists R_{j'}\ni j}\sum_{A\ni x}\frac{\|V_A\|}{\Delta_0} \\
&\le w\frac{\|V\|_\star}{\Delta_0},
\end{split}
\label{LAXTV}
\end{equation}
where $w$ is an order-one integer that depends only on the interaction range of $H_0$ and the geometry of the lattice:
\begin{equation}
w=\max_{j\in\Lambda}\left|\bigcup_{R_{j'}\ni j}R_{j'}\right|.
\label{w}
\end{equation}
For example, if $H_0$ consists of on-site potentials (i.e., $|R_j|=1$ for all $j\in\Lambda$), then $w=1$ in any spatial dimension \cite{Bravyi2011}; if $H_0$ is defined on a 1D lattice and only involves nearest-neighbor interactions (i.e., $|R_j|=2$ for all $j\in\Lambda$), we have $w=3$. In the following, we will focus on the large-$\Delta_0$ regime. This means $\Delta_0$ is large compared to $\|V\|_\star$, which sets a natural time scale and is considered to be of order one. In this regime, not only $\|T\|_\star$ (see Eq.~(\ref{LAXTV})) but also $\|V'\|_\star$ can be shown (see Appendix~\ref{bVp}) to be of the order of $\Delta_0^{-1}$. We will see that the error growth for a local observable can be bounded by these local interaction strengths.

\subsection{Lieb-Robinson bound} 
We introduce another crucial ingredient for deriving the error bound --- the Lieb-Robinson bound \cite{Lieb1972,Nachtergaele2006,Bravyi2006}. For our purpose, we consider a general setting where the many-body Hamiltonian $H(t)=\sum_{A\subseteq\Lambda}H_A(t)$ is time-dependent and quasi-local, with $\|H(t)\|_\mu$ uniformly bounded by some time-independent (but still $\mu$-dependent) constant. The many-body dynamics starting from time $t'$ is thus determined by
\begin{equation}
i\partial_tU(t,t')=H(t)U(t,t'),\;\;\;\;U(t',t')=1,
\end{equation}
which has a formal solution $U(t,t')\equiv\overrightarrow{{\rm T}}e^{-i\int^t_{t'}dsH(s)}$ ($\overrightarrow{{\rm T}}$: time ordering). Similar to the time-independent case \cite{Lieb1972}, we have an emergent ``soft" light cone for operator spreading, as is captured by the following theorem:
\begin{theorem}[Lieb-Robinson bound] %for time-dependent Hamiltonian] 
\label{LRThm}
Given a quasi-local time-dependent many-body Hamiltonian $H(t)$ with well-defined $\|H(t)\|_\mu<\infty$ for all $ \mu<\mu^*$ and $\forall t\in\mathbb{R}$, then for two arbitrary local operators $O_X$ and $O_Y$ and any $\kappa,\eta>0$ with $\kappa+\eta<\mu^*$, we have 
\begin{equation} 
\begin{split}
&\|[U(t,t')^\dag O_XU(t,t'),O_Y]\| \\
\le&2\min\{|X|,|Y|\}\|O_X\|\|O_Y\| \\
\times&e^{-\kappa[{\rm dist}(X,Y)-\frac{2C_d}{\kappa}(\frac{d}{e\eta})^d\int^t_{t'}ds\|H(s)\|_{\kappa+\eta}]},
\end{split}
\label{tLRB}
\end{equation}
where $U(t,t')$ is the unitary time-evolution operator generated by $H(t)$ during $[t',t]$ and $C_d$ is a constant (determined solely by the lattice geometry) that validates $|A|\le C_d (l_A+1)^d$ for all $ A\subseteq\Lambda$ \footnote{Note that when $|A|=1$ we have $l_A=0$ by definition. This explains why $l_A+1$ instead of $l_A$ appears in $|A|\le C_d(l_A+1)^d$.}.
\end{theorem}
Here by soft, we mean that there can be a tiny leakage from the light cone, i.e., the commutator does not rigorously vanish but only decays exponentially outside the light cone, as indicated by Eq.~(\ref{tLRB}). The proof of this theorem can be found in Appendix~\ref{PtLR}, which generalizes the proof for the time-independent case in Ref.~\cite{Hastings2010} in a rather straightforward way.

An important implication of Theorem~\ref{LRThm} is the following: for $H=H_0+V$ with $H_0$ being a commuting local Hamiltonian, as is the case of our setup (\ref{H0V}), the Lieb-Robinson bound of $H$ is essentially determined by the interaction strength of $V$, no matter how large the commuting interaction in $H_0$ is. This is because 
\begin{equation}
\begin{split}
&\|[O_X(t),O_Y]\| \\
=&\|[e^{iHt}e^{-iH_0t}e^{iH_0t}O_Xe^{-iH_0t}e^{iH_0t}e^{-iHt},O_Y]\| \\
=&\|[L(t)^\dag e^{iH_0t}O_Xe^{-iH_0t}L(t),O_Y]\|,
\end{split}
\end{equation}
where the Loschmidt operator $L(t)$ is given by
\begin{equation}
L(t)=e^{iH_0t}e^{-iHt}. 
\label{LE2}
\end{equation}
We emphasize that such a definition (\ref{LE2}) is only applicable throughout the present subsection and should not be confused with that in Eq.~(\ref{Lt}). One can check that $L(t)$ satisfies 
\begin{equation}
i\frac{d}{dt}L(t)=V^{H_0}(t)L(t),\;\;\;\;V^{H_0}(t)\equiv e^{iH_0t}Ve^{-iH_0t}.
\end{equation}
Since $H_0$ is commuting and local, denoting $l_0$ as the largest diameter of a local term in $H_0$, the diameter of $e^{iH_0t}V_Ae^{-iH_0t}$ in $V^{H_0}(t)$ should be no more than $l_A+2 l_0$ and its support should be no larger than $R_A$ (\ref{RA}), implying
\begin{equation}
\begin{split}
%&
\|V^{H_0}(t)\|_\mu %\\
&\le\max_{j\in\Lambda}\sum_{R_A\ni j}%\sum_{A\subseteq\Lambda}
\|V_A\| e^{\mu(l_A+2 l_0)} \\
\le &we^{2\mu  l_0}\|V\|_\mu.
\end{split}
\end{equation}
That is, the local interaction strength in $V^{H_0}(t)$ is bounded by a time-independent quantity that does not rely on the energy scale of $H_0$. Applying Eq.~(\ref{tLRB}) to $V^{H_0}(t)$ gives 
\begin{equation}
\begin{split}
&\|[O_X(t),O_Y]\| \\
\le& 2e^{\kappa l_0}\min\{|X|,|Y|\}\|O_X\|\|O_Y\| \\
\times&e^{-\kappa[{\rm dist}(X,Y)-\frac{2wC_d e^\eta}{\kappa}(\frac{d}{e\eta})^de^{2(\kappa +\eta) l_0}\|V\|_{\kappa +\eta} t]},
\end{split}
\label{HLR}
\end{equation}
where the additional prefactor $e^{\kappa l_0}$ comes from 
\begin{equation}
{\rm dist}(X_0,Y)\ge {\rm dist}(X,Y)-l_0,
\end{equation}
with $X_0$ being the support for $e^{iH_0t}O_Xe^{-iH_0t}$. Note that $|X|$ does not need to be replaced by $|X_0|$, because we can equivalently time evolve $O_Y$ to obtain a prefactor $\min\{|X|,|Y_0|\}$ with $Y_0$ being the support for $e^{-iH_0t}O_Ye^{iH_0t}$. Since $|X_0|\ge |X|$ and $|Y_0|\ge |Y|$, the optimal prefactor turns out to be $\min\{\min\{|X_0|,|Y|\},\min\{|X|,|Y_0|\}\}=\min\{|X|,|Y|\}$.

\begin{figure}[t]
    \centering
    \includegraphics[width=0.90\linewidth]{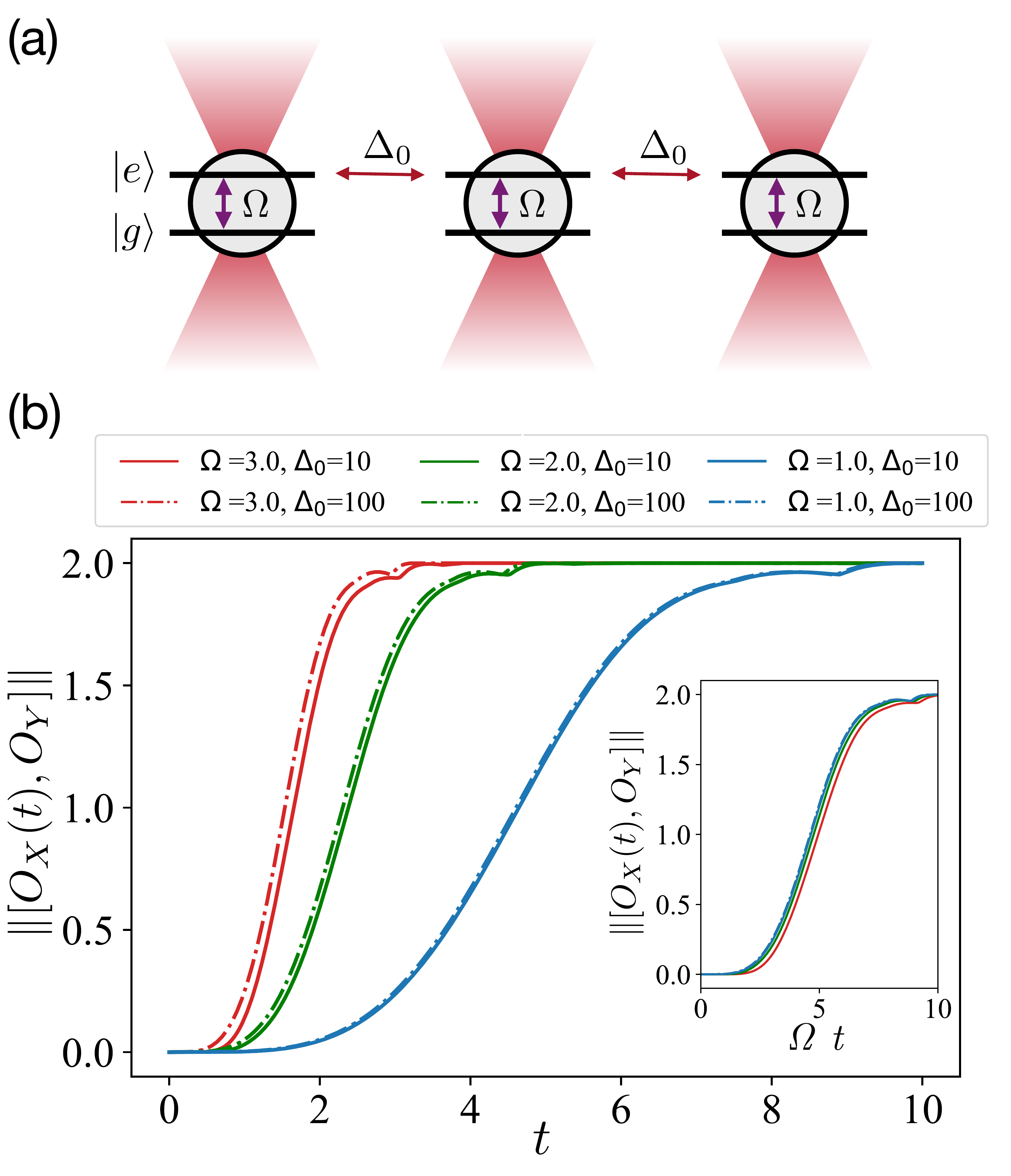}
    \caption{(a)~Schematic illustration of the parent Hamiltonian of the PXP model defined in Eq.~\eqref{eqn:pxp_def}. (b) Growth of $\|[O_X(t), O_Y]\|$ with the inset showing the collapse by rescaling the time as $\|V\|_\star t$. 
    The spreading of the operator depends on the {\it local} interaction strength $\|V\|_\star(=\frac{\Omega}{2})$ while the amplitude of the commuting interaction $\Delta_0$ barely affects the result as long as the gap is sufficiently large. 
    The operators considered here are chosen as $O_X = \sigma_{j=1}^y$ and $O_Y = \sigma_{j=6}^y$ where the system size is $N=10$. 
    The red, green, and blue lines denote $\Omega=1.0, 2.0, 3.0$, respectively, and the solid and chained lines correspond to $\Delta_0=10$ and 100, respectively.}
    \label{fig:figV4}
\end{figure}

\begin{figure}
    \centering
    \includegraphics[width=0.95\linewidth]{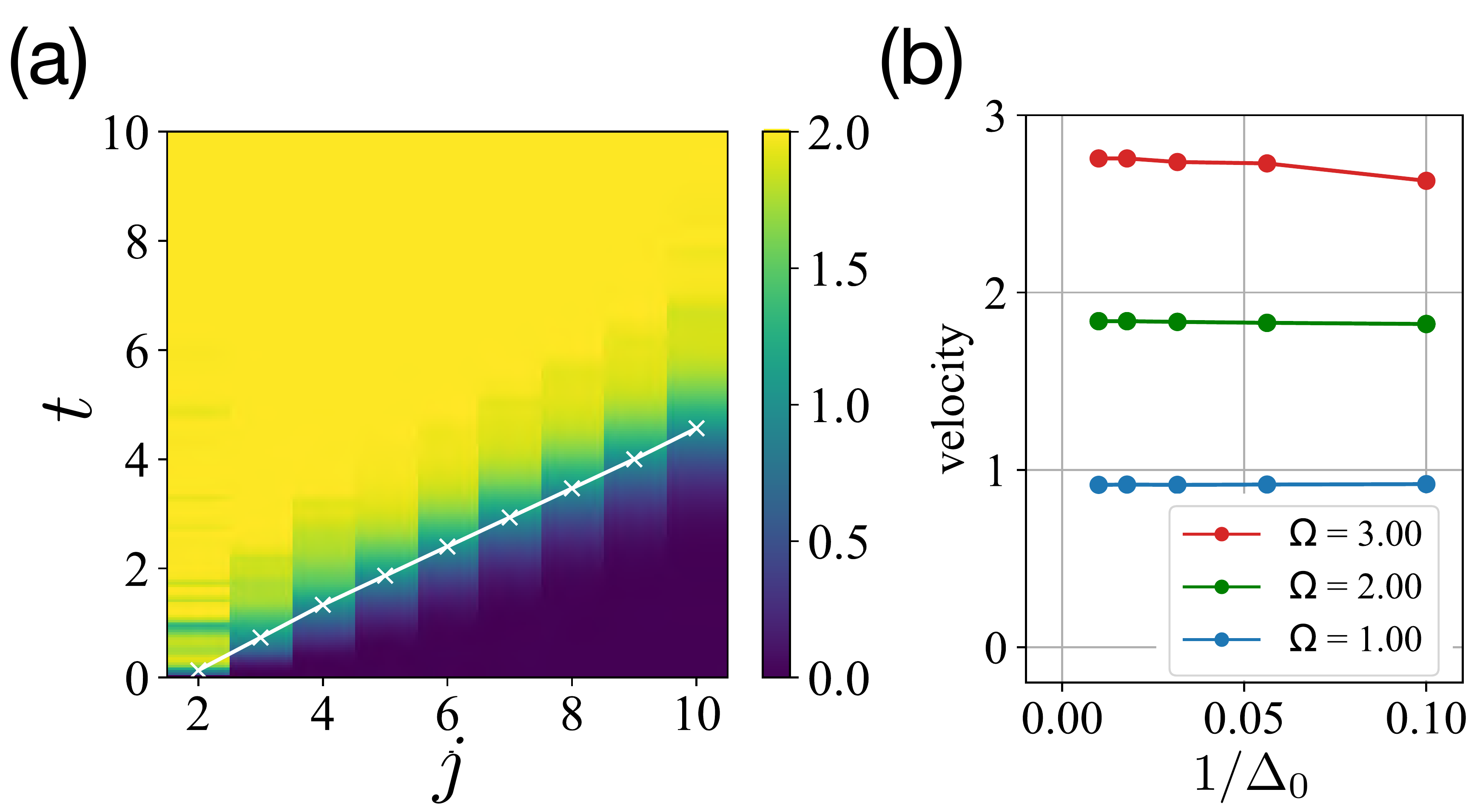}
    \caption{
    (a)~The linear light-cone structure of $\|[O_X(t),O_Y]\|$ where $O_X=\sigma_{j=1}^y$ and $O_Y=\sigma_{j}^y$ for the model defined in Eq.~\eqref{eqn:pxp_def}. 
    The white crossings denote the times when the threshold $(=1)$ is exceeded and the white lines are guide to eyes. 
    Here, the parameters of the Hamiltonian are set as $\Delta_0=10$ and $\Omega=2$.
    (b)~The independence of the operator-spreading velocity from the commuting interaction $\Delta_0$. The velocity is defined from the slope of the line consisting of the points where the threshold is exceeded (such as the white crosses in panel (a)). 
    %The small deviation at smaller $\Delta_0$ region for $\Omega=3$ is due to the insufficient gap. 
    In both panels, the system size is taken as $N=10$.}
    \label{fig:figV5}
\end{figure}

To demonstrate our findings, we consider the dynamics in the parent Hamiltonian of the PXP model defined on a chain with length $N$ under the open boundary condition~\cite{Turner2018,Turner2018b,Ho2019,Choi2019}:
\begin{equation}\label{eqn:pxp_def}
H_0=\frac{\Delta_0}{4}\sum_{j=1}^{N-1}(\sigma^z_j+1)(\sigma^z_{j+1}+1),\;\;\;\;
V=\frac{\Omega}{2}\sum_j\sigma^x_j,
%V=\Omega\sum_j\sigma^x_j,
\end{equation}
for which a graphical illustration is provided in Fig.~\ref{fig:figV4}(a).
The projector onto the degenerate low-energy manifold $\mathcal{H}_P$ is given as 
\begin{equation}
    P= \prod_j P_j = \prod_{j=1}^{N-1} \left [1- \frac{1}{4}(1+\sigma_j^z)(1+\sigma_{j+1}^z)\right],
    \label{PXPP}
\end{equation}
which prohibits adjacent excitations.
We show in Fig.~\ref{fig:figV4}(b) that, the operator spreading indeed relies on $\|V\|_\star$ but is barely affected by $\Delta_0$, which is implied from the Lieb-Robinson bound. This is also quantitatively confirmed from the velocity of the spreading (see Fig.~\ref{fig:figV5}). Note that while the Lieb-Robinson bound itself depends only on the interaction $V$, it is physically natural and yet consistent with the unsaturated bound that the velocity be modified when  the gap size becomes smaller. Such a behaviour is observed, e.g., at smaller $\Delta_0$ region for $\Omega=3$ as in Fig.~\ref{fig:figV5}(b).

\subsection{Error bound} %from the Lieb-Robinson bound}
We are now well-prepared to derive the error bound for quantum many-body systems. To make full use of locality, we focus on the error of a normalized \emph{local} observable $O_X$ supported on a finite region $X$. We first note that $H_1$ in Eq.~(\ref{HS}), which is explicitly given by
\begin{equation}
H_1\equiv H_0+\sum_{A\subseteq\Lambda}\mathscr{D}_A(V_A), %(block-diagonal component). 
\label{locH1}
\end{equation}
is block diagonalized and thus satisfies
\begin{equation}
PH_1=H_1P=PHP.
\label{PH1}
\end{equation}
This is because for the diagonal component $\mathscr{D}_A(V_A)$ of each local term $V_A$, we have
\begin{equation}
\begin{split}
&P\mathscr{D}_A(V_A)=PP_AV_AP_A \\
=&\left(\prod_{R_j\cap A=\emptyset} P_j\right)\left(\prod_{R_j\cap A\neq\emptyset} P_j\right)P_AV_AP_A \\
=&\left(\prod_{R_j\cap A=\emptyset} P_j\right)P_AV_AP_A=P_AV_AP_A\left(\prod_{R_j\cap A=\emptyset} P_j\right) \\
=&P_AV_AP_A \left(\prod_{R_j\cap A\neq\emptyset} P_j\right)\left(\prod_{R_j\cap A=\emptyset} P_j\right)\\
=&P_AV_AP_AP=\mathscr{D}_A(V_A)P,
\end{split}
\end{equation}
where we have used $[P_AV_AP_A,\prod_{R_j\cap A=\emptyset}P_j]=0$ (due to Eq.~(\ref{HHPP}) and the zero overlap between the supports of $V_A$ and $\prod_{R_j\cap A=\emptyset}P_j$), $P_A^2=P_A$ and $PP_A=P_AP=P\Leftrightarrow PQ_A=Q_AP=0$. It follows that
\begin{equation}
P\mathscr{D}_A(V_A)=\mathscr{D}_A(V_A)P=P\mathscr{D}_A(V_A)P=PV_AP,
\end{equation}
the sum of which gives rise to Eq.~(\ref{PH1}). This relation means that a local projection followed by a global projection is equivalent to a single global projection. We note that the two widely used expressions of the PXP Hamiltonian in the literature: $H_{\rm loc}=\frac{1}{4}\sum_j(1-\sigma^z_{j-1})\sigma^x_j(1-\sigma^z_{j+1})$ \cite{Turner2018,Turner2018b} and $H_{\rm glo}=P\sum_j\sigma^x_jP$ ($P$ is given in Eq.~(\ref{PXPP})) \cite{Ho2019,Choi2019} correspond to Eqs.~(\ref{locH1}) and (\ref{PH1}), respectively. These two Hamiltonians are qualitatively different, in the sense that the former is local while the latter is highly nonlocal. However, as a result of Eq.~(\ref{PH1}), both Hamiltonians give rise to exactly the same constrained dynamics if we start from a state in $\mathcal{H}_P$.

Thanks to Eq.~(\ref{PH1}), we can still rewrite the error in Eq.~(\ref{dec}) by choosing $S$ and $H_1$ to be the local SWT and the locally projected Hamiltonian (\ref{locH1}), respectively. While all the operators are on the many-body level, we can still apply Eq.~(\ref{decineq}) to obtain
\begin{equation}
\begin{split}
\epsilon(t)\le&\|e^{T}O_Xe^{-T}-O_X\|+\|L(t)O_XL(t)^\dag-O_X \| \\
+&\| e^{T_{H_1}(t)}O_Xe^{-T_{H_1}(t)}-O_X\|,
\end{split}
\label{decmb}
\end{equation}
where $T$ is now the \emph{local} Schrieffer-Wolff generator given in Eq.~(\ref{Tloc}), $L(t)=e^{-iH_1t}e^{i(H_1+V')t}=\overrightarrow{{\rm T}}e^{i\int^t_0dt'e^{-iH_1t'}V'e^{iH_1t'}}$ is the Loschmidt operator and $T_{H_1}(t)=e^{-iH_1t}Te^{iH_1t}$ is the local Schrieffer-Wolff generator in the interaction picture with respect to $H_1$. Recalling the special case $\mu=0$ in Eq.~(\ref{qloc}): %Defining the many-body norm for a general quasi-local operator $V=\sum_{A\subseteq\Lambda}V_A$ as
\begin{equation}
\|V\|_\star=\|V\|_{\mu=0}\equiv\max_{j\in\Lambda}\left(\sum_{A\ni j}\|V_A\|\right),
\end{equation}
we can bound the first term in Eq.~(\ref{decmb}) as
\begin{equation}
\begin{split}
&\|e^{T}O_Xe^{-T}-O_X\| \\
\le&\int_0^1 d\lambda \|e^{\lambda T}[T,O_X]e^{-\lambda T}\|=\|[T,O_X]\| \\
\le&\sum_{A\cap X\neq\emptyset}\|[T_A,O_X]\|\le\sum_{x\in X}\sum_{A\ni x}2\|T_A\| \\
\le&2|X|\|T\|_\star.
\end{split}
\label{TOXT}
\end{equation}
As for the remaining two terms, it is argued in Ref.~\cite{Gong2020a} on the basis of the light-cone picture that they should be of the order of $\frac{\|V\|_\star}{\Delta_0}$ and grow no faster than polynomially in time. In the following, we briefly review the argument and translate it into a rigorous result by using the Lieb-Robinson bound.

Let us first consider the rightmost term in Eq.~(\ref{decmb}). Once we succeed in bounding this term by a quantity that is polynomially large in time, we can also derive a polynomial bound on the middle term. We start with transferring the time dependence in the SWT generator onto the local observable $O_X$:
\begin{equation}
\begin{split}
&\| e^{T_{H_1}(t)}O_Xe^{-T_{H_1}(t)}-O_X\| \\
\le&\|[T_{H_1}(t),O_X]\|=\|[T,O^{H_1}_X(t)]\|,
\end{split}
\end{equation}
where $O^{H_1}_X(t)\equiv e^{iH_1t}O_Xe^{-iH_1t}$ is the observable in the Heisenberg picture. While the local interaction of $H_1$ can be very large for a large gap, we have proved in the previous subsection that the Lieb-Robinson velocity $v$ is essentially determined by the local interaction in $V$ and thus stays finite even in the infinite gap limit (see Eq.~(\ref{HLR}) and Fig.~\ref{fig:figV5}). Accordingly, denoting the radius of $X$ as $r_X$, that of time evolved $O^{H_1}_X(t)$ should effectively be $r_X+vt$. By effectively, we mean that there are exponentially decaying corrections outside the light cone. If we ignore these corrections, we immediately obtain a bound of the order of $\|T\|_\star (r_X+vt)^d$. If we carefully take into account these corrections, the result turns out to be qualitatively unchanged, as captured by the following theorem:
\begin{theorem}
\label{LRcom}
Suppose that the quantum dynamics generated by a many-body Hamiltonian defined on a $d$-dimensional lattice $\Lambda$ satisfies the Lieb-Robinson bound for any two local operators $O_X$ and $O_Y$:
\begin{equation}
\begin{split}
\|[O_X(t),O_Y]\|\le &2C\min\{|X|,|Y|\}\|O_X\|\|O_Y\| \\
\times &e^{-\kappa[{\rm dist}(X,Y)-vt]},
\end{split}
\label{CLR}
\end{equation}
where $C$, $\kappa$ and $v$ do not rely on the choices of $O_X$ and $O_Y$. Then given a quasi-local interaction $T=\sum_{A\subseteq\Lambda} T_A$, we have
\begin{equation}
\|[O_X(t),T]\|\le \|O_X\|\|T\|_\star p_\kappa(vt+r_X+\kappa^{-1}\ln(C|X|)),
\label{OXtT}
\end{equation}
where $r_X$ is the radius of $X$ defined as 
\begin{equation}
r_X\equiv\min\{r: \exists\;x\in\Lambda\;{\rm s.t.}\;{\rm dist}(x,x')\le r\;\forall x'\in X\},
\label{rXdef}
\end{equation}
and $p_\kappa(r)$ is a $\kappa$-parametrized polynomial of $r$ with degree $d$ given by
\begin{equation}
p_\kappa(r)=2(e^\kappa-1)b\left(r-\frac{d}{d\kappa}\right)(e^\kappa-1)^{-1}.
\end{equation}
Here $b(r)$ is the volume of a radius-$r$ ball, which is a monotonically increasing polynomial with degree $d$. 
\end{theorem}
The main idea to prove this theorem is to first bound $\|[O_X(t),T]\|$ by $\sum_{A\subseteq\Lambda}\|[O_X(t),T_A]\|$ and then treat the commutators in two separate ways depending on whether $A$ overlaps with the light cone. If there is nonzero overlap, we bound $\|[O_X(t),T_A]\|$ by $2\|O_X(t)\|\|T_A\|=2\|O_X\|\|T_A\|$. This part corresponds to the rough estimation in the above argument. Otherwise, for those $A$'s outside the light cone, we bound the commutator by the Lieb-Robinson bound in Eq.~(\ref{CLR}). It turns out that this part only effectively extends the light cone by an order-$\kappa^{-1}$ quantity, leaving the rough estimation qualitatively correct. A detailed proof of Theorem~\ref{LRcom} is available in Appendix~\ref{PLRc}.

\begin{figure}
\begin{center}
\includegraphics[width=7.5cm, clip]{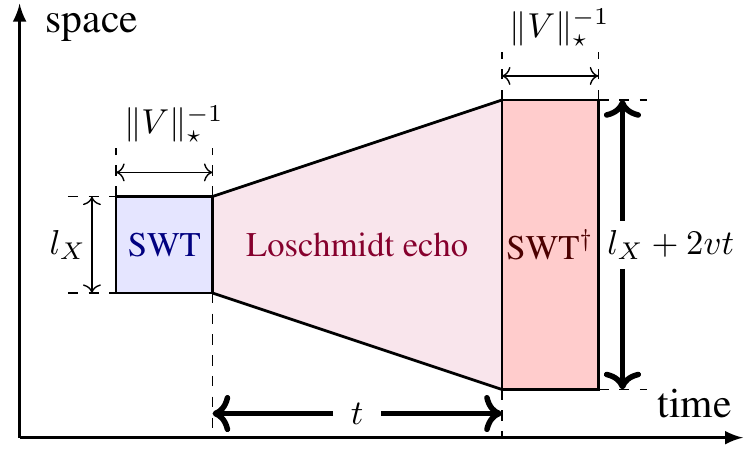}
       \end{center}
   \caption{Schematic illustration of error production in quantum many-body systems. The error is upper bounded by the \emph{space-time volume} of the operator spreading multiplied by $\mathcal{O}(\frac{\|V\|_\star^2}{\Delta_0})$ (see the analysis in the main text, especially Eqs.~(\ref{TOXT}) and (\ref{LtOX})). The dominant time and length scales (when $t\gg\|V\|_\star^{-1}$) are indicated by the thick arrows. The blue, purple and red regions correspond to the three terms on the right-hand side of Eq.~(\ref{decmb}) in order, which arise from the initial SWT, the Loschmidt echo and the inverse SWT in the interacting picture, respectively. Here only the projection onto a specific spatial direction is shown, so the entire volume should be obtained by taking power $d$ (spatial dimension) for the length followed by a time integral. According to the light-cone picture of operator spreading (numerically demonstrated in Fig.~\ref{fig:figV5}(a)), the volume and thus the error grow no faster than polynomially in time. }
   \label{figV6}
\end{figure}

Having the above analysis in mind, we are ready to bound the middle term in Eq.~(\ref{decmb}). We first note that $L(t)$ satisfies
\begin{equation}
\begin{split}
&\frac{d}{dt}L(t)=ie^{-iH_1t}V' e^{iH_1t}L(t)\;\;\;\; \\
\Leftrightarrow\;\;\;\;&i\frac{d}{dt}L(t)^\dag=L(t)^\dag e^{-iH_1t}V' e^{iH_1t},
\end{split}
\end{equation}
leading to
\begin{equation}
\begin{split}
&\|L(t)O_XL(t)^\dag-O_X\| \\
=&\|O_X-L(t)^\dag O_XL(t)\| \\
\le&\int_0^tdt'\|[e^{-iH_1t'}V'e^{iH_1t'},O_X]\| \\
=&\int_0^tdt'\|[V',O^{H_1}_X(t')]\|.
\end{split}
\label{LtOX}
\end{equation}
Since $V'$ is a quasi-local operator with bounded $\|V'\|_\star$ (see Appendix~\ref{bVp}), we can apply Theorem~\ref{LRcom} to obtain
\begin{equation}
\|[V',O^{H_1}_X(t)]\|\le \|V'\|_\star \tilde p(vt),
\end{equation}
where $\tilde p(r)\equiv p_\kappa(r+r_X+l_0+\kappa^{-1}\ln|X|)$ is a simplified notation of a degree-$d$ polynomial. Here we have used $\|O_X\|=1$ by assumption and explicitly set $C=e^{\kappa l_0}$ according to Eq.~(\ref{HLR}), with $l_0$ being the largest diameter of a local term in $H_0$. Defining $\tilde P(r)\equiv\int^r_0 dr' \tilde p(r')$ as a polynomial with degree $d+1$, we have
\begin{equation}
\|L(t)O_XL(t)^\dag-O_X\|\le v^{-1} \|V'\|_\star \tilde P(vt).
\end{equation}
Therefore, the overall error (\ref{decmb}) can be bounded by a quantity that is polynomially large in time:
\begin{equation}
\epsilon(t)\le [2|X|+\tilde p(vt)]\|T\|_\star+v^{-1}\tilde P(vt)\|V'\|_\star.
\end{equation}
Provided that the gap is sufficiently large, it has been shown that $\|V'\|_\star$ can asymptotically be bounded by $\frac{4wu\|V\|_\star^2}{\Delta_0}$, where $u$ is the degree of locality of $T$ (see Appendix~\ref{bVp}). Recalling that $\|T\|_\star\le w\frac{\|V\|_\star}{\Delta_0}$ (see Eq.~(\ref{LAXTV})), we have
\begin{equation}
\epsilon(t)\lesssim \frac{\|V\|_\star}{\Delta_0}w[2|X|+\tilde p(vt)+4uv^{-1}\|V\|_\star \tilde P(vt)],
\label{eptmb}
\end{equation}
which gives the desired main result on the error bound for quantum many-body systems. That is, the error is of the order of $\frac{\|V\|_\star}{\Delta_0}$ and grows no faster than a power law $t^{d+1}$. We summarize the above analysis in Fig.~\ref{figV6}, where we regard the (inverse) SWT as a time evolution governed by a Hamiltonian with $\mathcal{O}(\frac{\|V\|_\star^2}{\Delta_0})$ local-interaction strength for a duration $\|V\|_\star^{-1}$.

Finally, we remark that a sufficiently large $\Delta_0$ in Eq.~(\ref{eptmb}) implies a relatively long time before the saturation of the error to some constant no more than $2$. On an intermediately long time scale, the dominant term in the error bound is of the order of $\frac{\|V\|_\star^2v^d}{\Delta_0} t^{d+1}$. %reads $\frac{4uc_d\|V\|_\star^2v^d}{(d+1)\Delta}t^{d+1}$, where $c_d$ . 
This means that the constrained dynamics is a good approximation up to 
%\begin{equation}
$t^*\sim \left(\frac{\Delta_0}{v^d\|V\|_\star^2}\right)^{\frac{1}{d+1}}$.
%\end{equation} 

\section{Generalization to open quantum systems}\label{chapter4}
Remarkably, our idea to derive the error bound can also be generalized to few-level Markovian open quantum systems described by \emph{Lindblad master equations} \cite{Gorini1976, Lindblad1976}. This is motivated by the quantum Zeno effect \cite{Misra1977,Knight2000,Facchi2002} and is achieved by a non-unitary generalization of the SWT which differs from that in Ref.~\cite{Kessler2012}. Compared with a previous work concerning the error in quantum Zeno dynamics \cite{Zanardi2014}, our error bound has a more explicit form and is thus more physically comprehensible.

    \subsection{Setup and the main result}\label{opensmr}
        Let us consider the operator dynamics governed by the adjoint Lindblad equation
        \begin{align}
            \frac{d}{dt}O_t&=i[V, O_t]+\sum_j \qty(J_j^\dagger O_t J_j-\frac{1}{2}\qty{J_j^\dagger J_j, O_t})\notag\\
            \color{red}&=i(H_\mathrm{eff} O_t - O_t H_\mathrm{eff}^\dagger)+\sum_j J_j^\dagger O_t J_j\label{eq:adj_Lindblad_Schro_jump}\\
            & \equiv\mathscr{L}^\dagger[O_t],\label{eq:adj_Lindblad_Liouvillian}
        \end{align}
         where $ O_t $ is an arbitrary operator at time $ t $ in the Heisenberg picture, $ V $ is a Hermitian potential and $ J_j $'s are jump operators which are generally non-Hermitian. Moreover, as was done in Eq.~(\ref{eq:adj_Lindblad_Schro_jump}), we can decompose the dynamics into %the 
         two parts. That is, the first term is the Schr\"odinger evolution under the effective non-Hermitian Hamiltonian
        \begin{align}
        	H_\mathrm{eff}=V+\dfrac{i}{2}J_j^\dagger J_j=-i(-H_0+iV),
        \end{align}
        where
        \begin{align}
        	H_0\equiv \frac{1}{2}\sum_j J_j^\dagger J_j,
        \end{align}
        and the second term describes a stochastic quantum jump process. Such an equation of motion can be derived for a quantum system weakly coupled to a large environment \cite{Davies1974} or a continuously measured system upon averaging out the measurement outcomes \cite{Wiseman2010}. We assume that, in the absence of the unitary-dynamics part $ i[V, \;\cdot\;] $, the system has a \textit{decoherence-free subspace} (DFS)~\cite{Lidar1998}. By definition, the projector $ P $ onto the DFS satisfies $ J_j P=0 $ for all $ j$~\footnote{Rigorously speaking, a necessary and sufficient condition for generic decoherence-free dynamics in $ {\rm Span}[\{\ket{\phi_\alpha}\}_\alpha]$ is that all basis states $ \ket{\phi_\alpha} $ are degenerate eigenstates for all the jump operators $ \{J_j\}_j $: $ J_j\ket{\phi_\alpha} = c_j\ket{\phi_\alpha}$ for all $ \alpha $ and $ j $~\cite{Lidar1998}. This result comes from the definition property $J_jP_\alpha J_j^\dag-\frac{1}{2}\{J_j^\dag J_j,P_\alpha\}=0$ for all $j$ and $P_\alpha\equiv\ket{\phi_\alpha}\bra{\phi_\alpha}$.  However, when $ J_j $'s form a semisimple Lie algebra, these $ c_j $ become $ 0 $. This is the case for most examples including ours in  Sec.~\ref{sec:open_example}.} and can be thought as the projector onto the zero-energy manifold of $ H_0 $. In other words, there is no dynamics when we start from an arbitrary pure or mixed state in the DFS. We further assume that the zero-energy manifold of $ H_0 $ is gapped by $ \Delta_0 $ from the remaining subspace with non-zero energy.
        
Denoting $V_P\equiv PVP$ as the projected potential onto the DFS, we define the error of the constrained dynamics here as
        \begin{align}
            \epsilon(t)\equiv \norm{P e^{t\mathscr{L}^\dagger}[O]P-P e^{iV_P t}O e^{-iV_P t}P},
        \end{align}
        which is a natural generalization of Eq. (\ref{ept}). The main result in this section is the following upper bound:
        \begin{widetext}
            \begin{align}
                 \epsilon(t)
                %  &\le (e^{2\norm{T}}-1)\left\{1+e^{2\norm{T}}+e^{4\norm{T}}\qty[2\norm{V}+(e^{2\norm{T}}-1)\sum_j \norm{J_j}^2]t\right\}\\
                 &\le (e^{2\frac{\norm{V}}{\Delta_0}}-1)\left\{1+e^{2\frac{\norm{V}}{\Delta_0}}+e^{4\frac{\norm{V}}{\Delta_0}}\qty[2\norm{V}+(e^{2\frac{\norm{V}}{\Delta_0}}-1)\sum_j \norm{J_j}^2]t\right\}.\label{eq:error_bound_open}
             \end{align}
        \end{widetext}
        We expect that $ \norm{J_j}\sim \sqrt{\Delta_0} $, and thus there exists an order-one constant $ c $ such that $ \sum_j\norm{J_j}^2=c\Delta_0 $. Then, up to $\mathcal{O}(\frac{\norm{V}}{\Delta_0})$, we obtain from Eq.~(\ref{eq:error_bound_open}) the following asymptotic bound in the strong-dissipation limit $ \Delta_0\gg \norm{V} $: 
        \begin{align}
            \epsilon(t)\lesssim \dfrac{4\norm{V}}{\Delta_0}\qty[1+(1+c)\norm{V}t].\label{eq:open_bound_asymptotic}
        \end{align}
        
A few comments on Eq.~(\ref{eq:open_bound_asymptotic}) are in order. First, this bound can be understood as a quantitative manifestation of the quantum Zeno effect \cite{Misra1977,Knight2000,Facchi2002}. That is, rather counterintuitively, a coherent dynamics emerges in the limit of infinitely strong dissipation or measurement. Second, similarly to Eq. (\ref{eq:asymptotic_bound}), this bound implies the sudden jump and the linear growth of $ \epsilon(t) $. We will later demonstrate these behaviors in some simple models in Sec.~\ref{openex}. In addition, our bound is consistent with the semi-quantitative analysis in Ref.~\cite{Zanardi2014}, which essentially claims that the error is (at most) of the order of $\frac{\|V\|}{\Delta_0}$ on the time scale $t\sim\|V\|^{-1}$. Third, we note that this cannot be applied to an open system with infinite number of Lindblad operators, such as a dissipative lattice system \cite{Muller2012}. Analyzing the constrained dynamics or the quantum Zeno effect in open many-body systems \cite{Daley2014b} would be a very interesting, yet very challenging project for future studies.

      \subsection{Proof of the error bound for open quantum systems}%upper bound}
        The key to obtain the error bound for the open system is to consider the following \textit{non-unitary} SWT %Schrieffer-Wolff transformation
        \begin{align}
            S(-H_0+iV)S^{-1}=-H_0+iV_\mathrm{diag}+V'.
            \label{Heff}
        \end{align}
        Here, we take the \emph{Hermitian} operator $ T $ in the non-unitary SWT $ S=e^T $, which is actually Hermitian, to satisfy
        \begin{align}
            [T, H_0]=i(V-V_\mathrm{diag})=iV_\mathrm{off}.
        \end{align}
      Accordingly, the expression of $V'$ reads (cf. Eq.~(\ref{Vp}))
        \begin{equation}
            V'=\sum^\infty_{n=1} \frac{1}{n!}{\rm ad}^n_T\left(iV-\frac{1}{n+1}iV_{\rm off}\right).
            \label{iVp}
        \end{equation}
        One can see that the generator $ T $ here is related to that for a closed system just by multiplying $ i $, and therefore Inequalities (\ref{Tb}) and (\ref{Vpb}) are still valid in this case. It is worth mentioning that a different non-unitary SWT was developed in Ref.~\cite{Kessler2012} to block diagonalize the entire Lindbladian superoperator instead of the non-Hermitian effective Hamiltonian (\ref{Heff}). While the approach in Ref.~\cite{Kessler2012} is certainly useful for deriving the effective theory, it does not seem to be a convenient tool for bounding the error. This is because the validity of Eq.~(\ref{Tb}) requires the operators $ H_{0P} $ and $ H_{0Q} $ on the left-hand side of the Sylvester equation (\ref{TPQ}) to be Hermitian \cite{Bhatia1997}, which is not satisfied in the superoperator formalism \cite{Kessler2012}.
        
        Let us define $ O_t^S\equiv SO_t S $ and its equation of motion
        \begin{align}
            \dfrac{d}{dt}O_t^S=\mathscr{L}^{S\dagger}[O_t^S],
        \end{align}
        where $ \mathscr{L}^{S\dagger}[\;\cdot\;]=S\mathscr{L}^\dagger[S^{-1}\cdot S^{-1}]S $. Then, $ \epsilon(t) $ can be decomposed into three terms similarly to Eq. (\ref{decineq}):
        \begin{align}
        \begin{split}
             \epsilon(t)
            %  &\le \norm{P e^{t\mathscr{L}^{S\dagger}}[O]P-Pe^{iV_P t}Oe^{-iV_P t}P}\\
            % &\hspace{1em}+\norm{P e^{t\mathscr{L}^\dagger}[O]P-Pe^{t\mathscr{L}^{S\dagger}}[O]P}\\
            &\le \norm{ e^{t\mathscr{L}^\dagger}[O]-Se^{t\mathscr{L}^\dagger}[O]S}\\
            &\hspace{1em}+\norm{Pe^{-iV_{{\rm diag}} t} e^{t\mathscr{L}^{S\dagger}}[O]e^{iV_{{\rm diag}} t}P-POP}\\
             &\hspace{1em}+\norm{Se^{t\mathscr{L}^\dagger}[S^{-1}OS^{-1}]S-Se^{t\mathscr{L}^\dagger}[O]S}.
        \end{split}
        \end{align}
        This decomposition comes from the following: %way:
        \begin{align}
            \epsilon(t)&=\left\|Pe^{t\mathscr{L}^{S\dagger }}[O]P-P e^{iV_P t}O e^{-iV_P t}P\right.\notag\\
            &\hspace{2em}\left. +P e^{t\mathscr{L}^\dagger}[O]P-Pe^{t\mathscr{L}^{S\dagger}}[O]P \right\|\notag\\
            &\le \norm{Pe^{t\mathscr{L}^{S\dagger}}[O]P-P e^{iV_P t}O e^{-iV_P t}P}\notag\\
            &\hspace{1em}+\norm{P e^{t\mathscr{L}^\dagger}[O]P-PSe^{t\mathscr{L}^\dagger}[S^{-1}OS^{-1}]SP}\notag\\
            &\le\norm{Pe^{-iV_{{\rm diag}} t} e^{t\mathscr{L}^{S\dagger}}[O]e^{iV_{{\rm diag}} t}P-POP}\notag\\
            &\hspace{1em} +\left\| e^{t\mathscr{L}^\dagger}[O]-Se^{t\mathscr{L}^\dagger}[O]S\right.\notag\\
            &\hspace{1em}\left. +Se^{t\mathscr{L}^\dagger}[O]S-Se^{t\mathscr{L}^\dagger}[S^{-1}OS^{-1}]S \right\|\notag\\
            &\le \norm{Pe^{-iV_{{\rm diag}} t} e^{t\mathscr{L}^{S\dagger}}[O]e^{iV_{{\rm diag}} t}P-POP}\notag\\
            &\hspace{1em}+\norm{e^{t\mathscr{L}^\dagger}[O]-Se^{t\mathscr{L}^\dagger}[O]S}\notag\\
            &\hspace{1em}+\norm{Se^{t\mathscr{L}^\dagger}[O]S-Se^{t\mathscr{L}^\dagger}[S^{-1}OS^{-1}]S}.
        \end{align}
        Here we have also used the identity $ e^{iV_\mathrm{diag}t}P=Pe^{iV_\mathrm{diag}t}=Pe^{iV_Pt} $, which arises from $ PV_\mathrm{diag}=V_\mathrm{diag}P=V_P $.
        It is important to note that $ \norm{e^{t\mathscr{L}^\dagger}[O]}\le \norm{O}=1 $ \cite{tLO} and
        \begin{align}
        \begin{split}
             \norm{e^A O e^A - O} &\le \int_0^1 d\lambda\; \norm{e^{\lambda A}\{A, O\}e^{\lambda A}}\\
            &\le 2\norm{A}\int_0^1 d\lambda\; e^{2\lambda \norm{A}}\\
            &=e^{2\norm{A}}-1
        \end{split}
        \label{eAO}
        \end{align}
        for an arbitrary operator $ A $ to obtain the bound for the first and third terms:
        \begin{align}
        \begin{split}
             &\;\;\;\;\norm{ e^{t\mathscr{L}^\dagger}[O]-Se^{t\mathscr{L}^\dagger}[O]S} \\
             &+\norm{Se^{t\mathscr{L}^\dagger}[S^{-1}OS^{-1}]S-Se^{t\mathscr{L}^\dagger}[O]S}\\
            &\le\norm{e^{t\mathscr{L}^\dagger}[O]}(e^{2\norm{T}}-1)+e^{2\norm{T}}\norm{e^{t\mathscr{L}^\dagger}[S^{-1}OS^{-1}-O]}\\
            &\le e^{4\norm{T}}-1.\label{eq:open_bound_2_3_term}
        \end{split}
        \end{align}
        Let us then %consider to 
        obtain the bound for the second term. We define $ \tilde{O}_t\equiv e^{-iV_{{\rm d}} t} e^{t\mathscr{L}^{S\dagger}}[O]e^{iV_{{\rm d}} t} $ which satisfies the following equation of motion:
        \begin{align}
        \begin{split}
             \dfrac{d}{dt}\tilde{O}_t&=e^{-iV_{{\rm diag}} t}(-H_0+V')e^{iV_{{\rm diag}} t}\tilde{O}_t\\
             &+\tilde{O}_t e^{-iV_{{\rm diag}} t}(-H_0+V'^\dagger)e^{iV_{{\rm diag}} t}\\
             &+\sum_j e^{-iV_{{\rm diag}} t}S J_j^\dagger S^{-1}e^{iV_{{\rm d}} t}\tilde{O}_te^{-iV_{{\rm diag}} t}S^{-1} J_j Se^{iV_{{\rm diag}} t}.
        \end{split}
        \label{tilOeom}
        \end{align}
Note that $ \norm{\tilde{O}_t} $ itself is bounded by
        \begin{align}
        \begin{split}
            \norm{\tilde{O}_t}&=\norm{Se^{t\mathscr{L}^\dagger}[S^{-1}OS^{-1}]S} \\
            &\le e^{2\norm{T}}\norm{S^{-1}OS^{-1}}\le e^{4\norm{T}}.
         \end{split}
         \label{tilOn}
        \end{align}
        Using Eqs.~(\ref{tilOeom}) and (\ref{tilOn}) and the assumption $ J_j P=PJ_j^\dag=0 $, we obtain
        \begin{align}
        \begin{split}
            &\norm{Pe^{-iV_{{\rm d}} t} e^{t\mathscr{L}^{S\dagger}}[O]e^{iV_{{\rm d}} t}P-POP}=\norm{P\tilde{O}_tP-POP}\\
            &\le \int_0^t dt' \norm{P\dfrac{d}{dt'}\tilde{O}_{t'}P}\\
            &=\int_0^t dt' \left\|PV'e^{iV_{{\rm diag}} t'}\tilde{O}_{t'} e^{-iV_{{\rm diag}} t'}P\right.\\
            &\hspace{1em}+\left. Pe^{iV_{{\rm diag}} t'}\tilde{O}_{t'} e^{-iV_{{\rm diag}} t'}V'^\dagger P \right.\\
            &\left.+P\sum_j (SJ_j^\dagger S^{-1} -J_j^\dagger) e^{iV_{{\rm diag}} t}\tilde{O}_t e^{-iV_{{\rm diag}} t}(S^{-1}J_j S -J_j)P \right\|\\
            &\le \int_0^t dt' \norm{\tilde{O}_{t'}}\qty[2\norm{V'}+\sum_j \norm{J_j^\dagger}\norm{J_j}(e^{2\norm{T}}-1)^2]\\
            &\le e^{4\norm{T}}(e^{2\norm{T}}-1)\qty[2\norm{V}+(e^{2\norm{T}}-1)\sum_j \norm{J_j}^2]t\label{eq:open_bound_first_term}.
        \end{split}
        \end{align}
        Therefore, combining Eqs. (\ref{eq:open_bound_2_3_term}) and (\ref{eq:open_bound_first_term}), we obtain the bound for $ \epsilon(t) $ Eq. (\ref{eq:error_bound_open}).
        
Let us make a comment on the difference between the proof above and that for isolated systems in Sec.~\ref{pmr}. In the latter case, we perform the standard SWT for the full Hamiltonian, finding that the middle-step error production is due to a Loschmidt echo generated by an $\mathcal{O}(\frac{\|V\|^2}{\Delta_0})$ quantity and can thus be bounded by a time-linear term rather straightforwardly. In contrast, here we perform the SWT only for the anti-commutator part $-\frac{1}{2}\sum_j\{\;\cdot\;,J_j^\dag J_j \}=-\{\;\cdot\;,H_0\}$ in the Lindblad equation (\ref{eq:adj_Lindblad_Liouvillian}) in order to apply the rigorous norm inequalities, as mentioned in the beginning of the subsection. To handle the additional jump part $J_j^\dag(\;\cdot\;)J_j$, our strategy is to make full use of the property of DFS, i.e., $J_jP=PJ_j^\dag=0$, to modify the Schrieffer-Wolff transformed jump operator $S^{-1}J_jS$ in Eq.~(\ref{tilOeom}) into (see the fourth line in Eq.~(\ref{eq:open_bound_first_term}))
        \begin{equation}
        \tilde J_j\equiv S^{-1}J_jS-J_j. 
        \label{modjp}
        \end{equation}
        In this manner, the order of the modified jump operator is dramatically reduced from $\sqrt{\Delta_0}$ to $\|V\|/\sqrt{\Delta_0}$~\footnote{This reduction comes from $S^{-1}J_j S - J_j=\sum_{n=0}^{\infty} \frac{(-1)^n}{n!}{\rm ad}_T^nJ_j-J_j
            =-[T,J_j]+\text{higher-order terms}$ and Eq.~(\ref{Tb}). More precisely, we can prove $ \norm{S^{-1}J_jS-J_j}  \le \norm{J_j} (e^{2\norm{T} }-1)\simeq2\norm{J_j}\norm{T}$ following a similar procedure in Eq.~(\ref{eAO}).}. It is clear from Eqs.~(\ref{eq:open_bound_asymptotic}) and (\ref{b1}) that the only difference $c$ is contributed by these modified jump operators.
        
In particular, provided that $J_j=PJ_jQ$ for $\forall j$, which means each jump operator always sends a state outside the DFS into the DFS, the leading order of a modified jump operator in Eq.~(\ref{modjp}) constrained in the DFS, i.e., $P\tilde J_jP$, can be shown to be $J_jH_{0Q}^{-1}V_{QP}$ up to an unimportant phase factor. Regarding $H_{0Q}^{-1}$ as the Green's function outside the DFS at zero frequency, we can interpret $\tilde J_j$ as a coherent excitation from the DFS by $V$ followed by a transient propagation outside the DFS and finally a jump back to the DFS. This result is consistent with the standard second-order perturbation theory for adiabatic eliminations in open quantum systems \cite{Sorensen2012}.

    \subsection{Examples}\label{sec:open_example}
    \label{openex}
        To demonstrate the bound obtained above, here we give two simple examples. In fact, unlike the case of isolated quantum systems \cite{Gong2020a}, we have not obtained the \textit{worst} model for the error bound (\ref{eq:error_bound_open}). These two examples, however, qualitatively illustrate the possible error growth behaviors implied by our error bound, i.e., the initial sudden jump and the linear growth, respectively.
        
\begin{figure}
            \centering
            \includegraphics[width=1.0\linewidth]{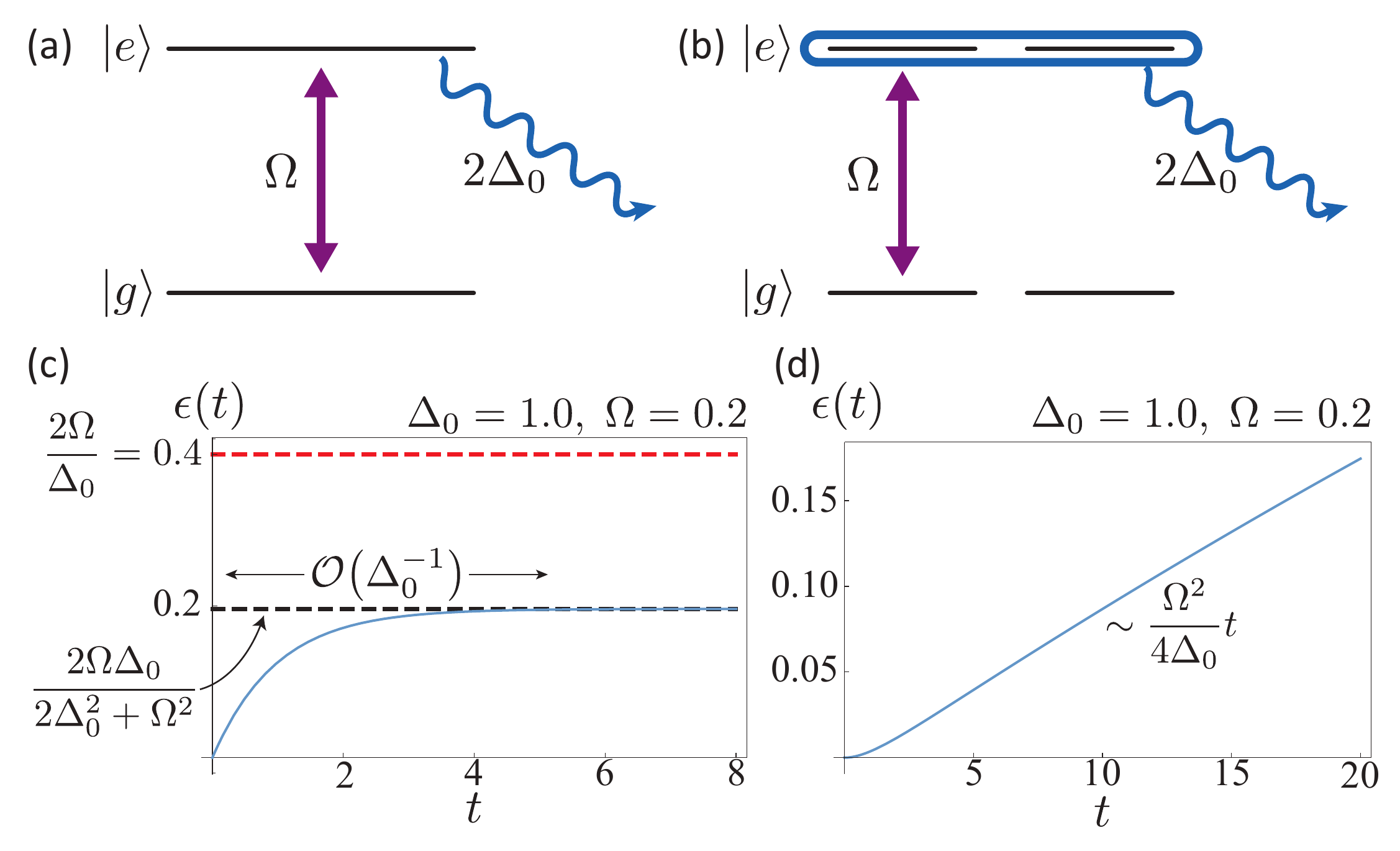}
            \caption{(a), (b) Schematic pictures for Examples 1 and 2 (see Eqs.~(\ref{openex1}) and (\ref{openex2})) and (c), (d) their errors. Each error shows either the sudden jump or the linear growth of $ \epsilon(t) $. (c) The error saturates to almost a half of the constant term (denoted as the red dashed line) of Eq. (\ref{eq:open_bound_asymptotic}) within $ t\sim \order{\Delta_0^{-1}} $. (d) The error grows linearly with a slope $ \sim\order{\Omega^2/\Delta_0}$.}
            \label{fig:figV7}
        \end{figure}

        \textit{Example 1.---}  We consider a two-level system with
        \begin{align}
        \begin{split}
             &J_1=\sqrt{2\Delta_0}\sigma^-=\sqrt{2\Delta_0}\ket{g}\bra{e},\\ 
             &V=\frac{\Omega}{2} \sigma^x,\;\; O=\sigma^y,
         \end{split}
         \label{openex1}
        \end{align}
        where $ \sigma^x $ and $ \sigma^y $ are the Pauli matrices. This model describes a resonantly driven two-level atom with an unstable excited state $|e\rangle$, the arguably simplest driven-dissipative system. It simply follows from Eq.~(\ref{openex1}) that $ H_0=\Delta_0 \ket{e}\bra{e} $ and $ P=\ket{g}\bra{g} $. For this example, one can obtain the analytical expression for the error $ \epsilon(t) $, which is a bit involved and thus not shown here. As shown in Fig.~\ref{fig:figV7}, the error saturates within time $ t\sim \order{\Delta_0^{-1}} $ to $ 2\Omega\Delta_0/(2\Delta_0^2+\Omega^2)\simeq \Omega/\Delta_0+\order{\Omega^2/\Delta_0^2} $, which is almost a half of the constant term of Eq. (\ref{eq:open_bound_asymptotic}) but already demonstrates a sudden jump.

        \medskip
        
        \textit{Example 2.---}  We consider two dissipatively coupled two-level systems with
        \begin{equation}
        \begin{split}
             J_1&=\sqrt{2\Delta_0}\sigma_1^-\sigma_2^-,\;\; V=\frac{\Omega}{2} \sigma_1^x,\\
             O&=\frac{1}{2}(\sigma_1^x\sigma_2^x+\sigma_1^y\sigma_2^y). 
             \end{split}
         \label{openex2}
        \end{equation}
This model describes two two-level atoms with only the first driven resonantly and only the double excited state being unstable. An equivalent model can be realized in a single four-level atom with states $|1\rangle$, $|2\rangle$, $|3\rangle$ and $|4\rangle$, where $|1\rangle$ ($|3\rangle$) is resonantly coupled to $|2\rangle$ ($|4\rangle$) with a Rabi frequency $\Omega$ and $|4\rangle$ undergoes spontaneous decay to $|1\rangle$ at a rate $2\Delta_0$. As for the original two-atom setting, one can easily check that $ H_0=\Delta_0\ket{ee}\bra{ee} $ and $ P=\ket{gg}\bra{gg}+\ket{ge}\bra{ge}+\ket{eg}\bra{eg} $, so the DFS excludes simultaneous excitations of two (adjacent) atoms, just like the constrained Hilbert space of the PXP model~\cite{Turner2018,Turner2018b}. The numerical result in Fig.~\ref{fig:figV7} (b) shows an almost linear growth of $ \epsilon(t) $, although a coefficient $ \sim\Omega^2/(4\Delta_0) $ is about 12 times as small as that of our obtained bound $ 4(1+c)\norm{V}^2/\Delta_0 $ (now $ c=2 $). Note that the modified jump operator constrained in the DFS is given by $P\tilde J_1P\simeq\frac{\Omega}{\sqrt{2\Delta_0}}|gg\rangle\langle ge|$. The operator dynamics generated by a single jump operator $P\tilde J_1P$ starting from $O$ in Eq.~(\ref{openex2}) simply gives $e^{-\frac{\Omega^2}{4\Delta_0}t}O$, which explains the slope observed in the numerical result since $\|e^{-\frac{\Omega^2}{4\Delta_0}t}O-O\|\simeq \frac{\Omega^2}{4\Delta_0}t$ for $t\ll\frac{\Delta_0}{\Omega^2}$.

\section{Conclusion and outlook}\label{chapter5}
Constrained quantum dynamics within a Hilbert subspace is a widely used approximation for gapped quantum systems, yet a quantitative and general justification has been lacking. Here, we have filled this gap by establishing a universal and rigorous error bound (see Theorem~\ref{Thm:ueb}). Our main idea is to decompose the error production into three steps on the basis of the SWT, and then bound the error generated in each step by an order $\frac{\|V\|}{\Delta_0}$ quantity. This result has been generalized to quantum many-body systems with local interactions by combining the local SWT with the Lieb-Robinson bound, and to open quantum systems using a non-unitary SWT. 

Our work rises a number of open questions. As already mentioned in Ref.~\cite{Gong2020a}, it stays unclear whether the slope and the intercept in the universal bound can simultaneously be saturated. In addition, the error bound for the single-state case and that for open quantum systems seems rather loose and it is worthwhile to think about how to tighten them. For further generalizations, one possibility is to consider the effect of higher-order SWTs \cite{Bravyi2011}. We believe that this can straightforwardly be done by following a similar three-step decomposition, and the error production from the Loschmidt echo is expected to be significantly suppressed by the higher order corrections. In particular, if $H_0$ has an isolated degenerate subspace $\mathcal{H}_P$ with eigenvalue $\omega_0$, the error arising from the Loschmidt echo is expected to be $\mathcal{O}(\frac{\|V\|^3}{\Delta_0^2}t)$ for the constrained dynamics within $\mathcal{H}_P$ generated by $H_P+\Sigma(\omega_0)$, where $\Sigma(\omega)=PVQ(\omega-H_Q)^{-1}QVP$ is the self-energy \cite{Feshbach1958,Feshbach1962,Brion2007}. Another possible generalization is to many-body open quantum systems \cite{Daley2014b}, as briefly mentioned in Sec.~\ref{opensmr}. This might be very challenging since there is some essential difference from isolated systems. For example, unlike the fact that commuting Hamiltonians include all the classical (spin) Hamiltonians, a classical Lindbladian is generally not a sum of mutually commutative superoperators. Once this generalization could be achieved, we would obtain a rigorous proof of the quantum many-body Zeno effect \cite{Zoller2014,Gong2017}. Extensions to long-range (power-law) interacting systems \cite{Hauke2013,Eisert2013,Gorshkov2014b,Lucas2019,Kuwahara2019} would also be an intriguing and rather urgent project, since long-range interactions appear naturally in many quantum simulators such as trapped ions \cite{Zhang2017b} and Rydberg atoms \cite{Bernien2017}. Last but not the least, we may consider possible improvement of the error bound for specific observables such as conserved quantities of $H_0$. This situation is relevant to quantum simulations of lattice gauge theories \cite{Halimeh2020}.

\acknowledgements
We acknowledge Takashi Mori for valuable comments. 
The numerical calculations were carried out with the help of QuTiP~\cite{qutip}.
Z.G. was supported by MEXT. 
N.Y. and R.H. were supported by Advanced Leading Graduate Course for Photon Science (ALPS) of Japan Society for the Promotion of Science (JSPS). N.Y. was supported by JSPS KAKENHI Grant-in-Aid for JSPS fellows Grant No. JP17J00743. N.S. acknowledges support of the Materials Education program for the future leaders in Research, Industry, and Technology (MERIT). R.H. was supported by JSPS KAKENHI Grant-in-Aid for JSPS fellows Grant No. JP17J03189).

\appendix

\section{Quasi-locality of $V'$ in the local Schrieffer-Wolff transformation}
\label{bVp}
In this Appendix, we prove that $V'$ in Eq.~(\ref{HS}) is a quasi-local interaction and that the strength is of the order of $\Delta_0^{-1}\|V\|_\star^2$, provided that $\Delta_0$ is large enough. We first introduce
\begin{equation}
\mathscr{O}_{\rm loc}(V)\equiv\sum_{A\subseteq\Lambda}\mathscr{O}_A(V_A),
\end{equation}
where $\mathscr{O}_A(V_A)\equiv P_AV_AQ_A+Q_AV_A P_A =V_A - \mathscr{D}_A(V_A)$. Following a procedure which is similar to the calculation for the global SWT, we obtain
\begin{equation}
V'=\sum^\infty_{n=1}\frac{1}{n!}{\rm ad}_T^n\left[V-\frac{1}{n+1}\mathscr{O}_{\rm loc}(V)\right].
\label{Vpn}
\end{equation}
Assuming that $V$ is \emph{$v$-local}, which means $V_A=0$ for $\forall A\subseteq\Lambda$ whenever $|A|>v$, and $H_0$ is $v_0$-local, then both $T$ (see Eq.~(\ref{Tloc})) and $\mathscr{O}_{\rm loc}(V)$ are at most $u$-local, with $u\ge\max\{v,v_0\}$ given by 
\begin{equation}
u\equiv\max_{|X|\le v,|Y|\le v_0}\left|\bigcup_{Y\cap X\neq\emptyset}Y\right|. 
\end{equation}
For example, for the parent Hamiltonian of the $PXP$ model, we have $v=1$, $v_0=2$ and $u=3$. 

In the following, we will bound each term in the series in Eq.~(\ref{Vpn}) and demonstrate the convergence for a sufficiently large $\Delta_0$. To this end, we need the following Lemma, which has essentially been pointed out in Ref.~\cite{Bravyi2011}:
\begin{lemma}
\label{vloc}
Given $V_1$ and $V_2$ as $v_1$-local and $v_2$-local interactions, $[V_1,V_2]$ is at most $(v_1+v_2-1)$-local and
\begin{equation}
\|[V_1,V_2]\|_\star\le2(v_1+v_2)\|V_1\|_\star\|V_2\|_\star
\label{V1V2}
\end{equation}
\end{lemma}
\emph{Proof.---} Since $[V_1,V_2]$ consists of the commutators of local terms in $V_1$ and $V_2$, which contain at most $v_1$ and $v_2$ sites, respectively, these commutators should contain at most $v_1+v_2-1$ sites. Here ``$-1$" comes from the fact that a nonvanishing commutator requires at least one-site overlap between the supports. To derive Eq.~(\ref{V1V2}), we only have to note that
%\begin{widetext}
\begin{equation}
\begin{split}
&\|[V_1,V_2]\|_\star \\
\le&\sum_{A\cup B\ni0,A\cap B\neq\emptyset}\|[V_{1,A},V_{2,B}]\|  \\
\le&2\left(\sum_{A\ni0,B\cap A\neq\emptyset}+\sum_{B\ni0,A\cap B\neq\emptyset}\right)\|V_{1,A}\|\|V_{2,B}\| \\
\le&2\sum_{A\ni0}|A|\|V_{1,A}\|\|V_2\|_\star+2\sum_{B\ni0}|B|\|V_1\|\|V_{2,B}\|_\star \\
\le& 2(v_1+v_2)\|V_1\|_\star\|V_2\|_\star.\;\;\;\;\square
\end{split}
\end{equation}
%\end{widetext}
We now turn to the bounding of the interaction norm $\|\cdot\|_\star$ of the $n$th term in Eq.~(\ref{Vpn}). First, similar to the operator norm, we have
\begin{equation}
\begin{split}
&\left\|V-\frac{1}{n+1}\mathscr{O}_{\rm loc}(V)\right\|_\star \\
\le&\frac{n}{n+1}\|V\|_\star+\frac{1}{n+1}\left\|\mathscr{D}_{\rm loc}(V)\right\|_\star\le w\|V\|_\star.
\end{split}
\label{VOV}
\end{equation}
where we have used $\|\mathscr{D}_{\rm loc}(V)\|_\star\le w\|V\|_\star$ (since the support of a local term $\mathscr{D}_A(V_A)$ is $R_A$ defined in Eq.~(\ref{RA})) and $w$ is defined in Eq.~(\ref{w}).
Recalling that $T$ and $\mathscr{O}_{\rm loc}(V)$ (and thus $V-\frac{1}{n+1}\mathscr{O}_{\rm loc}(V)$) are both at most $u$-local, using Lemma~\ref{vloc} and Eq.~(\ref{VOV}), we obtain
\begin{equation}
\begin{split}
&\left\|{\rm ad}_T^n\left[V-\frac{1}{n+1}\mathscr{O}_{\rm loc}(V)\right]\right\|_\star \\
\le& 2[u+nu-(n-1)]\|T\|_\star\left\|{\rm ad}_T^{n-1}\left[V-\frac{1}{n+1}\mathscr{O}_{\rm loc}(V)\right]\right\|_\star \\
\le&...\le w\|V\|_\star(2\|T\|_\star)^n\prod^n_{m=1}[(m+1)u-(m-1)],
\end{split}
\end{equation}
which implies
\begin{equation}
\begin{split}
&\left\|{\rm ad}_T^n\left[V-\frac{1}{n+1}\mathscr{O}_{\rm loc}(V)\right]\right\|_\mu \\
\le&\left\|{\rm ad}_T^n\left[V-\frac{1}{n+1}\mathscr{O}_{\rm loc}(V)\right]\right\|_\star e^{\mu(n+1)(u-1)}\\
\le&we^{\mu (u-1)}\|V\|_\star[2(u-1)e^{\mu(u-1)}\|T\|_\star]^n \\
&\times\prod^{n-1}_{m=0}\left[m+\frac{2u}{u-1}\right],
\end{split}
\label{admu}
\end{equation}
where we have used the fact that the diameter of a connected region with $(n+1)u-n$ sites is at most $(n+1)(u-1)$. Substituting Eq.~(\ref{admu}) into Eq.~(\ref{Vpn}) gives
\begin{equation}
\begin{split}
&\|V'\|_\mu \\
\le&we^{\mu (u-1)}\|V\|_\star \sum^\infty_{n=1}\frac{[2(u-1)e^{\mu(u-1)}\|T\|_\star]^n}{n!} \\
&\times\prod^{n-1}_{m=0}\left(m+\frac{2u}{u-1}\right)\\
=&w\|V\|_\star e^{\mu (u-1)}\{[1-2(u-1)e^{\mu(u-1)}\|T\|_\star]^{-\frac{2u}{u-1}}-1\},
\end{split}
\label{Vpfb}
\end{equation}
provided that $2(u-1)e^{\mu(u-1)}\|T\|_\star<1$, a sufficient condition for which is 
\begin{equation}
\Delta_0>2w(u-1)e^{\mu(u-1)}\|V\|_\star. 
\end{equation}
Here we have used the Taylor expansion 
\begin{equation}
(1-x)^{-\alpha}=\sum^\infty_{n=0}\alpha(\alpha+1)...(\alpha+n-1)\frac{x^n}{n!}. 
\end{equation}
When $\|T\|_\star$ ($\Delta_0$) is small (large) enough, Eq.~(\ref{Vpfb}) will be dominated by $4wu\|V\|_\star\|T\|_\star\sim \Delta_0^{-1}\|V\|_\star^2$. This can be considered as a many-body version of the Stark shift for a single atom, which is of the order of $\Delta^{-1}\Omega^2$ with $\Omega$ and $\Delta$ being the Rabi frequency and the detuning, respectively.

\section{Proof of Theorem~\ref{LRThm}}
\label{PtLR}
Given a finite region $X\subset\Lambda$, we introduce
\begin{equation}
I_X(t)\equiv\sum_{A\cap X\neq\emptyset}H_A(t).
\label{IX}
\end{equation}
Following the derivation in Ref.~\cite{Hastings2010}, given two local operators $O_X$ and $O_Y$ supported on finite regions $X$ and $Y$, we have
\begin{widetext}
\begin{equation}
\begin{split}
&\|[U(t+\epsilon,t')^\dag O_XU(t+\epsilon,t'),O_Y]\| \\
=&\|[O_X+i\epsilon[H(t),O_X],U(t,t')O_YU(t,t')^\dag]\|+\mathcal{O}(\epsilon^2) \\
=&\|[O_X+i\epsilon[I_X(t),O_X],U(t,t')O_YU(t,t')^\dag]\|+\mathcal{O}(\epsilon^2) \\
=&\|[e^{i\epsilon I_X(t)}O_Xe^{-i\epsilon I_X(t)},U(t,t')O_YU(t,t')^\dag]\|+\mathcal{O}(\epsilon^2)\\
=&\|[O_X,e^{-i\epsilon I_X(t)}U(t,t')O_YU(t,t')^\dag e^{i\epsilon I_X(t)}]\|+\mathcal{O}(\epsilon^2)\\
=&\|[O_X,U(t,t')O_YU(t,t')^\dag-i\epsilon [I_X(t),U(t,t')O_YU(t,t')^\dag]]\| 
+\mathcal{O}(\epsilon^2)\\
\le &\|[O_X,U(t,t')O_YU(t,t')^\dag]\| 
+2\epsilon\|O_X\|\|[I_X(t),U(t,t')O_YU(t,t')^\dag]\|+\mathcal{O}(\epsilon^2)\\
=&\|[U(t,t')^\dag O_XU(t,t'),O_Y]\| 
+2\epsilon\|O_X\|\|[U(t,t')^\dag I_X(t)U(t,t'),O_Y]\|+\mathcal{O}(\epsilon^2).
%&\|[U(t+\epsilon,t')^\dag O_XU(t+\epsilon,t'),O_Y]\| \\
%=&\|[O_X+i\epsilon[H(t),O_X],U(t,t')O_YU(t,t')^\dag]\|+\mathcal{O}(\epsilon^2) \\
%=&\|[O_X+i\epsilon[I_X(t),O_X],U(t,t')O_YU(t,t')^\dag]\|+\mathcal{O}(\epsilon^2) \\
%=&\|[e^{i\epsilon I_X(t)}O_Xe^{-i\epsilon I_X(t)},U(t,t')O_YU(t,t')^\dag]\|+\mathcal{O}(\epsilon^2)\\
%=&\|[O_X,e^{-i\epsilon I_X(t)}U(t,t')O_YU(t,t')^\dag e^{i\epsilon I_X(t)}]\|+\mathcal{O}(\epsilon^2)\\
%=&\|[O_X,U(t,t')O_YU(t,t')^\dag-i\epsilon [I_X(t),U(t,t')O_YU(t,t')^\dag]]\| \\
%&+\mathcal{O}(\epsilon^2)\\
%\le &\|[O_X,U(t,t')O_YU(t,t')^\dag]\| \\
%+&2\epsilon\|O_X\|\|[I_X(t),U(t,t')O_YU(t,t')^\dag]\|+\mathcal{O}(\epsilon^2)\\
%=&\|[U(t,t')^\dag O_XU(t,t'),O_Y]\| \\
%+&2\epsilon\|O_X\|\|[U(t,t')^\dag I_X(t)U(t,t'),O_Y]\|+\mathcal{O}(\epsilon^2).
\end{split}
\end{equation}
\end{widetext}
%Integrating $t$ from $t'$ to $t$ 
Replacing $t$ by $s$ followed by integrating it from $t'$ to $t$ and applying Eq.~(\ref{IX}), we obtain
\begin{equation}
\begin{split}
&\|[U(t,t')^\dag O_XU(t,t'),O_Y]\| \\
\le&\|[O_X,O_Y]\|+2\|O_X\| \\
\times&\sum_{A\cap X\neq\emptyset}\int^t_{t'} ds\|[U(s,t')^\dag H_A(s)U(s,t'),O_Y]\|.
\end{split}
\label{UOXU}
\end{equation}
By further defining 
\begin{equation}
C_{XY}(t,t')\equiv\max_{\|O_X\|=1}\|[U(t,t')^\dag O_XU(t,t'),O_Y]\|
\label{CXY}
\end{equation}
for a given $O_Y$ and noting that there is no special requirement for Eq.~(\ref{UOXU}) being valid, we must have
\begin{equation}
\begin{split}
C_{XY}(t,t')\le &C_{XY}(t',t') \\
+&2\sum_{A\cap X\neq\emptyset}\int^t_{t'}ds \|H_A(s)\| C_{AY}(s,t'),
\end{split}
\label{CXYineq}
\end{equation}
where $C_{XY}(t',t')=0$ if $X\cap Y=\emptyset$ and otherwise is upper bounded by $2\|O_Y\|$.

\begin{figure}[b]
    \includegraphics[width=7cm, clip]{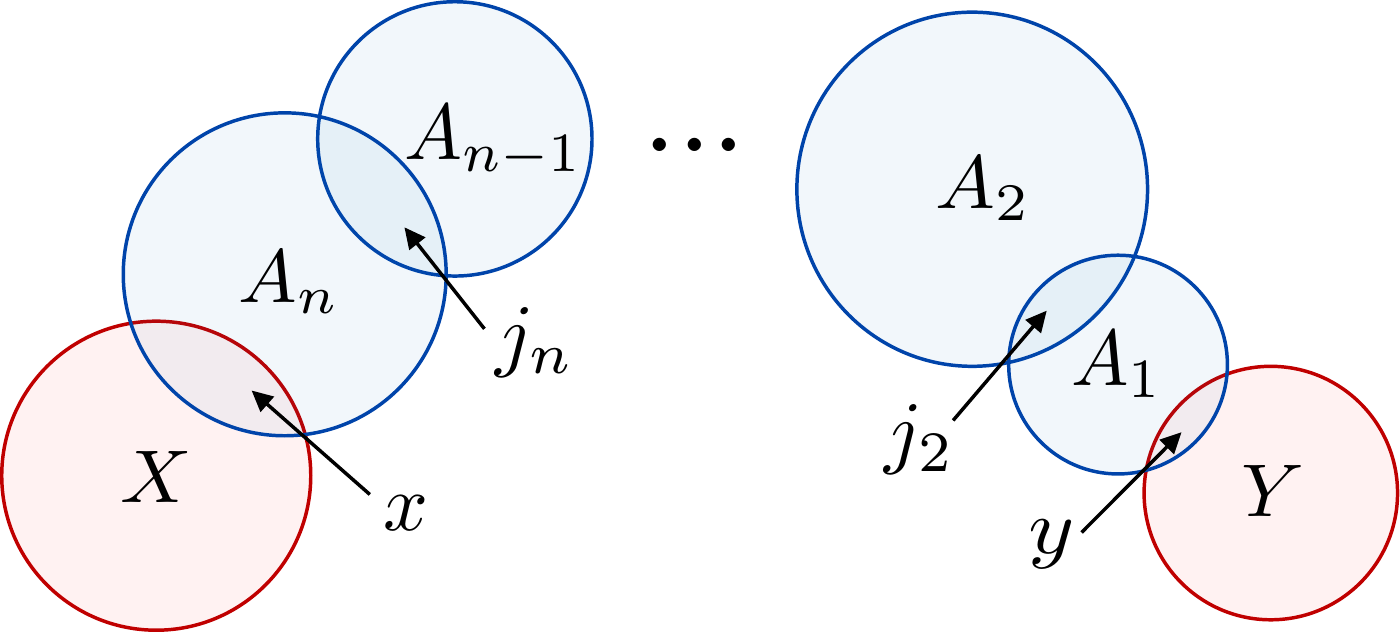}
    \caption{Schematic illustration of the sums in Eqs.~(\ref{jeder}) and (\ref{ander}). For a set of regions $\{A_j\}^n_{j=1}$ satisfying $X\cap A_n\neq\emptyset$, $A_m\cap A_{m-1}\neq\emptyset$ ($m=2,3,...,n$) and $A_1\cap Y\neq\emptyset$, we can always find $x\in X\cap A_n$, $j_m\in A_m\cap A_{m-1}$ and $y\in A_1\cap Y$ such that ${\rm dist}(X,j_n)\le{\rm dist}(x,j_n)\le l_{A_n}$, ${\rm dist}(j_m,j_{m-1})\le l_{A_m}$ and ${\rm dist}(Y,j_2)\le{\rm dist}(y,j_2)\le l_{A_1}$. This explains the first ``$\le$" in Eq.~(\ref{jeder}) as well as the alternative bound in Eq.~(\ref{ander}).}
    \label{fig:figV8}
\end{figure}

Applying Eq.~(\ref{CXYineq}) to itself iteratively, we obtain a series on the right-hand side:
\begin{widetext}
\begin{equation}
C_{XY}(t,t')\le2\|O_Y\|\sum^\infty_{n=0}2^n\sum_{X\cap A_n\neq\emptyset}\sum_{A_n\cap A_{n-1}\neq\emptyset}...\sum_{A_1\cap Y\neq\emptyset}\int^t_{t'}ds_n\int^{s_n}_{t'}ds_{n-1}...\int^{s_2}_{t'}ds_1\prod^n_{m=1}\|H_{A_m}(s_m)\|.
\end{equation}
\end{widetext}
%Since each term in the sum is non-negative and in this case
The non-negative integrand in each term can be bounded by
\begin{widetext}
\begin{equation}
\begin{split}
&\sum_{X\cap A_n\neq\emptyset}\sum_{A_n\cap A_{n-1}\neq\emptyset}...\sum_{A_1\cap Y\neq\emptyset}\prod^n_{m=1}\|H_{A_m}(s_m)\| \\
\le& \sum_{x\in X}\sum_{A_n\ni x}\sum_{j_n\in A_n}\sum_{A_{n-1}\ni j_n}...\sum_{A_1\ni j_2} e^{\kappa[l_{A_n}-{\rm dist}(x,j_n)+l_{A_{n-1}}-{\rm dist}(j_n,j_{n-1})+...+l_{A_1}-{\rm dist}(j_2,Y)]}\prod^n_{m=1}\|H_{A_m}(s_m)\| \\
\le & \sum_{x\in X}\sum_{A_n\ni x}\sum_{j_n\in A_n}\sum_{A_{n-1}\ni j_n}...\sum_{A_1\ni j_2} e^{-\kappa{\rm dist}(X,Y)}\prod^n_{m=1}\|H_{A_m}(s_m)\|e^{\kappa l_{A_m}} \\
\le & |X|e^{-\kappa{\rm dist}(X,Y)}\prod^n_{m=1}\left(\max_{j\in\Lambda}\sum_{A_m\ni j}|A_m|\|H_{A_m}(s_m)\|e^{\kappa l_{A_m}}\right)
\le |X|e^{-\kappa{\rm dist}(X,Y)}C_d^ne^{n\eta}\left(\frac{d}{e\eta}\right)^{nd}\prod^n_{m=1}\|H(s_m)\|_{\kappa+\eta},
\end{split}
\label{jeder}
\end{equation}
\end{widetext}
where we have used the triangular inequality for ${\rm dist}$ and
\begin{equation}
|A|\le C_d (l_A+1)^d\le C_d\left(\frac{d}{e\eta}\right)^d e^{\eta (l_A+1)},\;\;\forall\eta>0.
\end{equation}
Note that the prefactor $|X|$ in Eq.~(\ref{jeder}) can be replaced by $|Y|$ if we bound the first line differently by taking the weighted sum (see Fig.~\ref{fig:figV8})
\begin{equation}
\sum_{y\in Y}\sum_{A_1\ni y}\sum_{j_2\in A_1}\sum_{A_2\ni j_2}...\sum_{A_n\ni j_n}e^{\kappa\sum^n_{m=1}[l_{A_m}-{\rm dist}(j_{m+1},j_m)]},
\label{ander}
\end{equation}
where $j_1\equiv y$ and $j_{n+1}\equiv X$. After summing up the series and using $\|[U(t,t')^\dag O_XU(t,t'),O_Y]\|\le \|O_X\|C_{XY}(t,t')$, we obtain the expected Lieb-Robinson bound (\ref{tLRB}) for unitary dynamics generated by a time-dependent Hamiltonian.

\section{Proof of Theorem~\ref{LRcom}}
\label{PLRc}
%Without loss of generality, we assume that $X$ is covered by a ball centered at $x$ with radius $r_X$. It then follows
According to the definition of $r_X$ in Eq.~(\ref{rXdef}), we can always find a fixed site $x\in\Lambda$ such that
\begin{equation}
{\rm dist}(A,x)\le{\rm dist}(A,X)+r_X,\;\;\;\;\forall A\subseteq\Lambda.
\label{AxAX}
\end{equation}
This is because ${\rm dist}(A,B^{r_X}_x)\le{\rm dist}(A,X)$ due to $X\subseteq B^{r_X}_x\equiv\{x'\in\Lambda:{\rm dist}(x',x)\le r_X\}$ and 
\begin{equation}
\begin{split}
{\rm dist}(A,B^{r_X}_x)+r_X \ge&{\rm dist}(a^*,x^*)+r_X \\
\ge&{\rm dist}(a^*,x^*)+{\rm dist}(x^*,x) \\
\ge&{\rm dist}(a^*,x)\ge{\rm dist}(A,x), 
\end{split}
\end{equation}
where $a^*\in A$ and $x^*\in X$ minimize $\{{\rm dist}(a,x):a\in A,x\in X\}$ to ${\rm dist}(A,X)$. Since $\|[O_X(t),O_Y]\|$ always has a trivial bound $2\|O_X\|\|O_Y\|$, we have 
\begin{equation}
\begin{split}
&\|[O_X(t),O_Y]\|\le 2\|O_X\|\|O_Y\| \\
&\times\min\{1,C\min\{|X|,|Y|\} e^{-\kappa [{\rm dist}(X,Y)-vt]}\},
\end{split}
\end{equation}
where $C$ is the constant in the assumption Eq.~(\ref{CLR}) such that the theorem applies to the case of Eq.~(\ref{HLR}) (with $C=e^{\kappa l_0}$). Therefore, we can bound $\|[O_X(t),T]\|$ by 
\begin{equation}
\begin{split}
&\|[O_X(t),T]\|\le\sum_{A\subseteq\Lambda}\|[O_X(t),T_A]\| \\
\le&2\|O_X\|\sum_{A\subseteq\Lambda}\|T_A\|\min\{1,C|X|e^{-\kappa[{\rm dist}(A,X)-vt]}\} \\
\le&2\|O_X\|\sum_{A\subseteq\Lambda}\|T_A\|\min\{1,C|X|e^{-\kappa[{\rm dist}(A,x)-r_X-vt]}\} \\
=&2\|O_X\|\sum_{A\cap B_x^{r(t)}\neq\emptyset}\|T_A\| \\
+&2C\|O_X\|\sum_{A\cap B_x^{r(t)}=\emptyset}\|T_A\||X|e^{-\kappa[{\rm dist}(A,x)-r_X-vt]},
\end{split}
\label{TOXH1}
\end{equation}
where we have used Eq.~(\ref{AxAX}) and $r(t)$ is given by
\begin{equation}
r(t)=r_X+\lfloor vt+ \kappa^{-1}\ln (C|X|)\rfloor.
\label{rt}
\end{equation}
Applying the same technique used in the last step in Eq.~(\ref{TOXT}), the first term in Eq.~(\ref{TOXH1}) can be bounded by
\begin{equation}
\sum_{A\cap B_x^{r(t)}\neq\emptyset}2\|T_A\|\le 2|B^{r(t)}_x|\|T\|_\star,
\label{nleer}
\end{equation}
where the volume of a ball with radius $r$, which should be a strictly increasing function of $r$, can generally be expressed as a degree-$d$ polynomial:
\begin{equation}
|B_x^r|=\sum^d_{m=0} c_m r^m\equiv b(r).
\label{VBr}
\end{equation}
The coefficients in Eq.~(\ref{VBr}) depend on the geometry of the lattice (nevertheless, $c_0=1$ in any case) and $d$ is the spatial dimension. For example, for a 2D triangular lattice we have $b(r)=1+3r+3r^2$; for a 3D cubic lattice we have $b(r)=1+\frac{8}{3}r+2r^2+\frac{4}{3}r^3$.

\begin{figure}
\begin{center}
\includegraphics[width=7.5cm, clip]{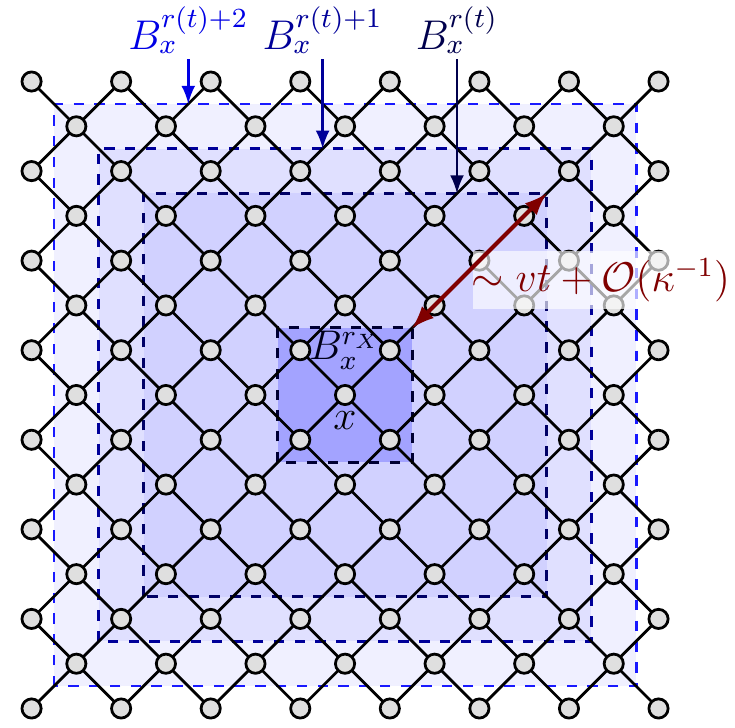}
       \end{center}
   \caption{ Schematic illustration of the decomposition technique for the case of square lattice. Here $B_x^{r_X}$ is the minimal ball that fully covers $X$, the support of $O_X$. At time $t$, we bound the commutator between $O_X(t)$ and the operators $T_A$'s whose supports have overlap with $B^{r(t)}_x$ using $2\|O_X\|\|T_A\|$. Here $r(t)\sim vt + \mathcal{O}(\kappa^{-1})$, where $v$ is the Lieb-Robinson velocity and $\kappa^{-1}$ is the correlation length. Otherwise, if $A\cap B^{r(t)}=\emptyset$, we use the Lieb-Robinson bound with the distance chosen to be $r(t)+n$, where $n$ is the largest integer such that $A\cap B^{r(t)+n}=\emptyset$,. }
   \label{figV9}
\end{figure}

We move on to bound the sum in the last line of Eq.~(\ref{TOXH1}). All the sets with zero overlap with $B_x^{r(t)}$ can further be divided into two classes -- one with and the other without overlap with $B_x^{r(t)+1}$. For the latter class, we can again divide it into two classes with and without overlap with $B_x^{r(t)+2}$. This procedure can be repeated until the full lattice is covered (see Fig.~\ref{figV9}). Since for $\forall A\cap B_x^{r(t)+n}=\emptyset$ we have
\begin{equation}
\begin{split}
&{\rm dist}(A,x)\\
\ge& r(t)+n+1 \\
\ge& r_X+vt +\kappa^{-1}\ln (C|X|)+n,
\end{split}
\end{equation}
we can bound the sum by
\begin{equation}
\begin{split}
&\sum_{A\cap B_x^{r(t)}=\emptyset}2C\|T_A\||X|e^{-\kappa[{\rm dist}(A,x)-r_X-vt]} \\
&\le\sum^\infty_{n=0}\sum_{A\cap B_x^{r(t)+n}=\emptyset,A\cap B_x^{r(t)+n+1}\neq\emptyset} 2\|T_A\| e^{-\kappa n} \\
&\le\sum^\infty_{n=0}\sum_{A\cap( B_x^{r(t)+n+1}\backslash B_x^{r(t)+n} )\neq\emptyset}2\|T_A\| e^{-\kappa n} \\
&\le 2\|T\|_\star\sum^\infty_{n=0}(|B_x^{r(t)+n+1}|- |B_x^{r(t)+n}|)e^{-\kappa n}.
\end{split}
\label{leer}
\end{equation}
Combining Eqs.~(\ref{TOXH1}), (\ref{nleer}) and (\ref{leer}), we obtain
\begin{equation}
\begin{split}
&\|[O_X(t),T]\| \\
\le&2(e^\kappa-1)\|O_X\|\|T\|_\star\sum^\infty_{n=1}|B_x^{r(t)+n}|e^{-n\kappa}.
\end{split}
\label{mittel}
\end{equation}
In terms of the explicit polynomial expression $b(r)=\sum^d_{m=0}c_m r^m$ for $|B_x^r|$, the sum in Eq.~(\ref{mittel}) can be calculated to be
%\begin{equation}
%\begin{split}
%&\sum^\infty_{n=1}|B_x^{r(t)+n}|e^{-n\kappa} 
$\sum^\infty_{n=1}|B_x^{r(t)+n}|e^{-n\kappa}=\sum^\infty_{n=1}\sum^d_{m=0}c_m[r(t)+n]^me^{-n\kappa}
=\sum^\infty_{n=1}\sum^d_{m=0}c_m[r(t)-\frac{d}{d\kappa}]^me^{-n\kappa}%\\
=\sum^d_{m=0} c_m[r(t)-\frac{d}{d\kappa}]^m(e^\kappa-1)^{-1}
= b(r(t)-\frac{d}{d\kappa})(e^\kappa-1)^{-1}$.
%\sum^\infty_{n=1}|B_x^{r(t)+n}|e^{-n\kappa}=\sum^\infty_{n=1}\sum^d_{m=0}c_m[r(t)+n]^me^{-n\kappa}
%=\sum^\infty_{n=1}\sum^d_{m=0}c_m\left[r(t)-\frac{d}{d\kappa}\right]^me^{-n\kappa}\\
%=\sum^d_{m=0} c_m\left[r(t)-\frac{d}{d\kappa}\right]^m(e^\kappa-1)^{-1}
%= b\left(r(t)-\frac{d}{d\kappa}\right)(e^\kappa-1)^{-1}.
%\end{split}
%\end{equation}
Since $b(r)$ is a strictly increasing function of $r$, we can get rid of $\lfloor\;\cdot\;\rfloor$ in Eq.~(\ref{rt}) while leaving Eq.~(\ref{mittel}) valid, which now becomes the desired result in Eq.~(\ref{OXtT}). %which is a polynomial of $t$ with degree $d$. 
In particular, for $1$D systems, we have $b(r)=2r+1$ and thus
\begin{equation}
\begin{split}
&\|[O_X(t),T]\|\\
\le&2\|O_X\|\|T\|_\star\left[2r(t)+1+\frac{2e^\kappa}{e^\kappa-1}\right] \\
\le&2\|O_X\|\|T\|_\star\left[2vt+2r_X+2\kappa^{-1}\ln(C|X|)+\frac{3e^\kappa-1}{e^\kappa-1}\right].
\end{split}
\end{equation}

\ \\
\ \\
\ \\
\ \\
\bibliography{GZP_references}

\begin{thebibliography}{70}
\bibitem [{tLO()}]{tLO}%
  \BibitemOpen
  \href@noop {} {}\bibinfo {note} {This is because (i) $e^{\mathcal{L}^\dag
  t}[\;\cdot\;]$ is dual to a quantum channel and thus has a Kraus
  representation $\sum_\alpha K_\alpha^\dag \cdot K_\alpha$ with $\sum_\alpha
  K_\alpha^\dag K_\alpha=I$; (ii) according to the variational definition,
  $\|\cdot\|=\langle\psi|\cdot|\phi\rangle$ for two specific normalized states
  $|\psi\rangle$ and $|\phi\rangle$. Using the Cauchy-Schwarz inequality, we
  obtain $\|e^{\mathcal{L}^\dag t}[O]\|=\sum_\alpha \langle\psi|K_\alpha^\dag O
  K_\alpha|\phi\rangle\le \sum_\alpha \|\langle\psi|K_\alpha^\dag\| \|O\|
  \|K_\alpha|\phi\rangle\|\le \|O\|\sqrt{(\sum_\alpha \langle\psi|K_\alpha^\dag
  K_\alpha|\psi\rangle)(\sum_\alpha \langle\phi|K_\alpha^\dag
  K_\alpha|\phi\rangle)}=\|O\|$.}\BibitemShut {Stop}%
\end{thebibliography}

\end{document}